\documentclass[12pt]{article}

\usepackage{graphicx}
\usepackage{amsmath,amssymb}
\usepackage{graphics}
\usepackage{graphicx}
\usepackage[latin1]{inputenc}
\usepackage{epsfig}

\newcommand{\R}{\mathbb{R}}

\newcommand{\E}{\mathbb{E}}
\newcommand{\lambdavect}{\boldsymbol{\lambda}}

\newcommand{\pr}{\mathcal{P}}
\newcommand{\br}{\mathcal{B}}
\newcommand{\calH}{\mathcal{H}}

\begin{document}

\date{}

\title{\vspace{-0.8cm}\kern-2.0truecm{\Large\bf  An assets-liabilities dynamical  model of    banking system and  systemic risk governance}}


\author{\kern-0.0truecm
\normalsize{\bf Lorella Fatone} \\
{\kern-0.0truecm\small Dipartimento di Matematica}\\
{\kern-0.0truecm\small Universit\`a di Camerino}\\
{\kern-0.0truecm\small  Via Madonna delle Carceri 9, 62032
Camerino (MC),
Italy}\\
{\kern-0.0truecm\small Ph. n.+39-0737-402558, FAX
n.+39-0737-632525,
E-mail: lorella.fatone@unicam.it}\\[4mm]
\normalsize {\kern-0.0truecm\bf Francesca Mariani}\\
{\kern-0.0truecm\small  Dipartimento di Scienze Economiche e Sociali}\\
{\kern-0.0truecm\small Universit\`a Politecnica delle Marche}\\
{\kern-0.0truecm\small  Piazza Martelli 8, 60121 Ancona (AN), Italy}\\
{\kern-0.0truecm\small Ph. n.+39-071-2207243, FAX
n.+39-071-2207102,
E-mail: f.mariani@univpm.it}\\[4mm]
 }
\maketitle

\begin{abstract}
\noindent

We consider  the problem of governing  systemic risk in  an assets-liabilities  dynamical model of     banking system.
In the model  considered each  bank is  represented   by   its  assets and its     liabilities.  The capital reserves of a bank are the    difference between     assets and  liabilities of the bank.  A bank is solvent  when its  capital reserves  are greater or equal to zero  otherwise  the bank is failed. 
The  banking system dynamics  is defined by an   initial value problem    for  a system of    stochastic differential equations whose 
independent variable is time    and whose dependent variables are     the  assets and   the   liabilities of the banks.  The     banking system  model   presented  generalizes those   discussed  in 
\cite{Fouque1},  \cite{nostro1} and describes a  homogeneous population  of banks.
 The main features of    the model  are a      cooperation  mechanism among   banks and  the possibility  of the  (direct)  intervention of the monetary authority in the  
 banking system dynamics.
 We call systemic  risk or systemic event  in a   bounded  time interval   the fact  that  in that time interval  at least  a given fraction of the   banks     fails. 
The probability   of    systemic risk   in a  bounded time interval  is evaluated   using statistical simulation.   
The  systemic risk  governance pursues the  goal of  keeping the  probability  of  systemic risk  in a bounded time interval  between two given thresholds.
The monetary authority is responsible for the  systemic risk   governance.
The   governance consists in  the  choice  of  the assets and of  the  liabilities    of  a kind of  ``ideal bank''    as functions of time   and   in  the  choice     of  the rules   that regulate the cooperation mechanism among    banks. 
These rules    are obtained solving 
   an   optimal control problem  for  the  pseudo mean field  approximation of the   banking system  model. 
   The   governance  induces  the banks of  the   system   to behave like the  ``ideal bank''. 
Shocks acting on  the   assets or  on  the  liabilities of the banks are simulated. Numerical examples  of    systemic risk  governance in presence and in absence  of shocks acting on the banking system    are  studied. 
%

%
\end{abstract}
\date
\maketitle

\section{Introduction}\label{sec1}
%
The notion of systemic risk refers to the risk of a collapse  of an entire system rather than simply the failure of individual parts of it.
Systemic risk and systemic risk governance are  important research topics  that have applications in  many  different contexts such as, for example, physics,  biology, engineering,  finance.  
We limit our attention  to the  modeling of systemic risk in banking systems. For a survey of the  
use of mathematical models  in the   study  of  systemic risk in  a more general  context we refer to \cite{HandSR}  and to the references therein.

This paper  is concerned with   measurement,   monitoring and   governance  of systemic risk    in an assets-liabilities
   dynamical model of   banking system. Recently several dynamical models of  banking systems  based on  stochastic differential equations have been  studied,  see, for example,    \cite{Fouque1}, \cite{Fouque2},  \cite {Papan1}, \cite{nostro1}.
We  present a banking system model that generalizes those presented   in \cite{Fouque1},  \cite{nostro1} and exploits some ideas   taken from   \cite{Haldane}, \cite{May}, \cite{Petersen1}, \cite{Petersen2}.  
That is we consider a continuous-time  dynamical model of banking system  where  each bank holds assets and has liabilities  that are  stochastic  processes of time.
Assets and  liabilities of each bank as functions of time  are 
defined implicitly  by     an   initial value problem    for  a system of    stochastic differential equations.
The capital reserves of a bank are  defined as the difference between     assets and   liabilities of the bank.  A bank is solvent  when its  capital reserves  are greater or equal to zero  otherwise  the bank is failed. 
A    political/technical authority    is  responsible for the banking system management and, in particular,    is  responsible  for  the  systemic risk governance.
For convenience  we refer to this  authority  as       monetary authority.

The  model  proposed  describes a homogeneous population of banks where each bank interacts with the other banks and with the monetary authority. 
Note that the     homogeneity of the bank population   implies      that  all the  banks   of the model  behave in the same way. 
The main features of  the model        are  a cooperation mechanism among banks that regulates   the    inter-bank borrowing and lending activity and the  possibility of the  (direct) intervention of the monetary authority in the banking system dynamics. 
The  cooperation   mechanism  among  banks  (see \cite{Fouque1},  \cite{nostro1}) is  based on the idea that  ``who has more (assets, liabilities) gives to those who have less (assets, liabilities)''.    The intervention of the monetary authority in the banking system dynamics consists 
in  the  choice of   two  functions      representing   respectively  the assets and the liabilities of a kind of  ``ideal bank'' as functions of time and in  the  choice of    the rules   that regulate  the cooperation mechanism among    banks. 

In  the banking system model   proposed  realistic situations  of banking distress  due to the deterioration 
of the quality of the assets  and/or  of  the  liabilities of the banks can be modeled. 
Shocks that hit the banking system are simulated  with jumps in  the volatilities 
of  the stochastic differential equations satisfied by  the assets and by    the liabilities  of the banks 
and    with jumps  of    the correlation coefficients of the  stochastic differentials of the diffusion  terms   that appear on the right hand  side
  of   the assets and    of   the liabilities   equations.

We call systemic  risk or systemic event  in a   bounded  time interval   the fact  that  in that time interval  at least  a given fraction of the   banks  of the model     fails. 
Given  a banking system model  we use  statistical simulation  to evaluate  the probability   of    systemic risk   in a  bounded time interval. 
The  action of the cooperation mechanism among banks     produces a  reduction of   the    default  probability  of the individual  bank at the expenses of   an increment   of    the    default  probability    of all   or almost all the banks of the   banking system.  This last event is called   ``extreme'' systemic risk. 

When the number of banks  of the model   goes to infinity a   heuristic  approximation of the banking system model     called  ``pseudo mean field approximation''  is  introduced. 
This      approximation is inspired  to the mean field approximation of statistical mechanics and is based on the homogeneity of the bank population.
The pseudo mean field  approximation   is a stochastic   dynamical system  with two degree of freedom.

 A  method   to  govern    the probability of systemic risk  in a  bounded  time interval is presented. 
 The goal of the governance   is to keep the  probability  of systemic risk in a bounded time interval  between two given thresholds. 
The    governance     exploits  the choice made by the monetary authority of  the assets and of the liabilities      of  a kind of   ``ideal bank''   as functions of time  and    the solution of   a  stochastic  optimal control problem  for  the  pseudo mean field approximation of the   banking system  model.  
In fact   in a  homogeneous bank  population  when there are enough  banks,  
all the banks behave like  a kind of    ``mean bank'' and the   ``mean bank''   behaviour is   approximated    with the  behaviour of 
  the  pseudo mean field  approximation   of the banking system model. 
  This last  behaviour is  forced   to  be similar to the  behaviour of   the     ``ideal bank''   solving     a stochastic optimal control problem
  for the    pseudo mean field   approximation of  the banking system model. 
Thanks  to the homogeneity of the bank population,     the governance of   the pseudo mean field  approximation   
   is easily   translated   in the governance of   the entire bank population.
More specifically  it is  translated in  the  rules of the cooperation mechanism among banks. 
In this way the  systemic risk governance induces  the individual banks to behave as the ideal bank.
  Shocks  on  the   assets and on the  liabilities of the banks are simulated and
   numerical  examples  of    systemic risk  governance   in presence and in absence  of  shocks 
    are presented. 

 In the  scientific literature several  banking system models have been suggested.
 For example in   \cite{Fouque1},  \cite{Fouque2}, \cite{nostro1}    banking system models      consisting  in   initial value problems  for  systems of stochastic differential equations have been  studied. In    \cite{Fouque1},  \cite{Fouque2} the  dependent variables of the  stochastic 
 differential equations that define the model  are  the   log-monetary reserves of the banks as  functions of time and    there is  a cooperation mechanism that  regulates    the borrowing and lending activity among banks.      
 Moreover the probability of systemic risk in a bounded time interval       is  studied  using the mean field approximation and   the theory of large deviations. 
The model  presented  in  \cite{nostro1}   generalizes those  presented in  \cite{Fouque1},  \cite{Fouque2}. In particular in \cite{nostro1}  a model    with   two  cooperation mechanisms is studied. The first  cooperation mechanism regulates    the borrowing and lending activity among banks while  the second one  describes     the borrowing and lending activity between  banks  and monetary authority. 
Furthermore a  technique to govern the probability of systemic risk in a bounded time interval is   introduced and studied. 
In  \cite{Haldane},  \cite{May},  \cite{Petersen1}, \cite{Petersen2}    assets-liabilities models  of  banking systems  are presented.
 Each bank is modeled  by its assets and its liabilities. 
Time independent   (static)    \cite{Haldane},  \cite{May}  and time dependent   (dynamic) 
 \cite{Petersen1}, \cite{Petersen2}      assets-liabilities   banking system models have been studied. 
 In  \cite{Petersen1}, \cite{Petersen2}    the    assets and   the    liabilities  of the banks  are further decomposed  
  in the sum of  more specific   addenda  and   the time dynamics of  each addendum     is specified. 
Finally  in \cite{Haldane},  \cite{May}  the  analogies  between systemic risk in  banking systems  and systemic risk  in several other  domains  of     science and engineering are explored.
 


The paper is organized as follows.
In Section \ref{sec2}  an   assets-liabilities  banking system  model   is defined. 
%
In Section \ref{sec3} the definition of  systemic risk in a bounded time interval is given and the  implications of the  presence of the 
 cooperation mechanism among banks and 
  of  the homogeneity of the bank population
on the systemic risk probability  are  investigated.
In Section \ref{sec4}  the mean field and the  pseudo mean field   approximations of the  banking system  model 
defined  in Section \ref{sec2} are  discussed.
In Section \ref{sec5}   an optimal control problem for the pseudo mean field  approximation  of  the banking system   model is 
  solved and the optimal control  found is translated in the     rules that determine the   functioning of the cooperation mechanism  among  banks.  
  Finally in Section \ref{sec6}    a method   to govern  systemic risk in a bounded time interval   is presented  and  
  some numerical examples of systemic risk  governance of  banking systems in presence and in absence  of  shocks are discussed.

\section{The   banking  system model}\label{sec2}
%
%
  Let $t$ be a real variable that  denotes  time and  $N>1$ be   a positive integer representing  the number of banks present in the banking system model  at time $t=0$.   The superscript $i$ labels the   $i$-th bank,  $i=1,2,\ldots,N$. 
The activities of each bank are partitioned in  the following categories: interbank loans,   external assets, deposits  and interbank borrowings.
The assets of a bank are made of  the interbank loans and  of the external assets of the bank.
The liabilities of a bank are made of the deposits  and of   the interbank borrowings of the bank.
 The assets $a_{t}^{i}$  of the $i$-th bank at time $t \ge 0$   are the sum of    the interbank loans  $\iota_{t}^{i}$ at time   $t \ge 0$,  and   of  the
 external assets  $e_{t}^{i}$ at time   $t \ge 0$,  of the $i$-th   bank,    $ i=1,2,\ldots,N$, that is:
\begin{equation}\label{assets}
a_{t}^{i}=\iota_{t}^{i}+e_{t}^{i}, \quad t \ge 0, \qquad i=1,2,\ldots,N.
\end{equation}
 The  liabilities $l_{t}^{i}$  of the $i$-th bank at time $t \ge 0$ are  the sum of  the  deposits  $d_{t}^{i}$ at time   $t \ge 0$,  and  of the  interbank borrowings  $b_{t}^{i}$ at time   $t \ge 0$,  of the $i$-th  bank, $ i=1,2,\ldots,N$, that is:
\begin{equation}\label{liab}
l_{t}^{i}=d_{t}^{i}+b_{t}^{i}, \quad t \ge 0, \qquad i=1,2,\ldots,N.
\end{equation}
 The previous four categories of activities are balanced in the bank capital. The capital reserves  or ``net worth''  of the $i$-th bank,  $c_{t}^{i}$,  
    at time  $t \ge 0$,  are defined as  the difference between   assets $a_{t}^{i}$   at time $t \ge 0$  and liabilities $l_{t}^{i}$  at time $t \ge 0$,   of the $i$-th bank, $i=1,2,\ldots,N$, that is: 
\begin{equation}\label{reserves}
c_{t}^{i}=a_{t}^{i}-l_{t}^{i}, \quad t \ge 0, \qquad i=1,2,\ldots,N.
\end{equation}
A bank is solvent when  its assets are greater or equal to  its  liabilities, that  is  the $i$-th bank is solvent  at time   $t \ge 0$, if
\begin{equation}\label{solvency}
c_{t}^{i}=a_{t}^{i}-l_{t}^{i} \ge 0, \quad t \ge 0, \qquad i=1,2,\ldots,N.
\end{equation}
When    the capital reserves  $c_{t}^{i}$,  $t \ge 0$,  of the $i$-th bank become  negative for the first time    during the time evolution the  $i$-th bank is failed, $i=1,2,\ldots,N$. The failed banks are removed from the banking system model. 
%
Note that in the models studied in this paper  the   assets, the liabilities and  the capital reserves   of each bank  are  stochastic processes      
    of time,  in particular this means  that the inequality (\ref{solvency}) must be considered on each path of the  stochastic process that represents  the capital reserves. 
  That is  a bank  can be failed on a path of its  capital reserves  and  can be solvent on a different path 
of its  capital reserves.
Equations (\ref{assets}),  (\ref{liab}), (\ref{reserves})   are  a simple   model of bank capital, more  advanced   models  of bank capital   can be found, for example,  in  \cite{Diamond}.

In  \cite{Petersen1}, \cite{Petersen2}    the dynamics of each  addendum present  on the right hand side    of  (\ref{assets}),  (\ref{liab})  is  specified,  instead here  we  specify only  the dynamics  of  the assets  $a_{t}^{i}$, $t \ge 0$, and  of the liabilities $l_{t}^{i}$, $t \ge 0$,  $i=1,2,\ldots,N$.
In fact  we assume that    the assets  and the liabilities   of  the   banks   are    stochastic processes  of time   defined  implicitly by  the following   system of   stochastic differential equations:   
\begin{eqnarray}
&& d a_{t}^{i}= a_{t}^{i} \mu_{a}  dt + a_{t}^{i} \sigma_{a}  dW_{t}^{i}, \quad t>0, \,  i=1,2,\ldots,N , \label{al1} \\
&& d l_{t}^{i}= l_{t}^{i} \mu_{l}  dt + l_{t}^{i} \sigma_{l}  dZ_{t}^{i},\quad    t>0, \,  i=1,2,\ldots,N ,  \label{al2}
\end{eqnarray}
with the  initial conditions:
\begin{eqnarray}  \label{alCIgeneral}
&&a_{0}^{i}  =\tilde{a}_{0}^{i}, \quad l_{0}^{i}  =\tilde{l}_{0}^{i}, \quad i=1,2,\ldots,N,                
\end{eqnarray}
where  $\sigma_{a}=\sigma_{a,t}$,  $t>0$,  $\sigma_{l}=\sigma_{l,t}$,  $t>0$,  are piecewise  constant  positive  functions of time
 and $\mu_{a}$, $\mu_{l}$ are real constants. 
In (\ref{alCIgeneral})   $\tilde{a}_{0}^{i}$, $\tilde{l}_{0}^{i}$, $ i=1,2,\ldots,N,$  are random
variables that, for simplicity,    we assume  to be   concentrated in a point with
probability one. With abuse of notation  we use  the same symbols to    denote  the random variables
 and the points where the random variables are concentrated. 
We  assume  $\tilde{a}_{0}^{i}>0$, $\tilde{l}_{0}^{i}>0$, $\tilde{a}_{0}^{i}-\tilde{l}_{0}^{i}>0$, $ i=1,2,\ldots,N$, 
that is   we assume that at time $t=0$ all the banks are solvent with probability one.

The stochastic processes $W_{t}^{i},$ $Z_{t}^{i},$ $ t \ge 0$,   in (\ref{al1}), (\ref{al2}) are standard  Wiener processes,  such that $W_{0}^{i}=0,$ $Z_{0}^{i}=0,$  and  $dW_{t}^{i},$    $dZ_{t}^{i},$  $t>0$,  are their
stochastic differentials,  $ i=1,2,\ldots,N$.  We assume that:
\begin{eqnarray}\label{rho}
&&\E(dW_{t}^{i} dW_{t}^{j})=\rho_{a}^{2} \, dt,\,   \, \, i\ne j, \qquad  \E(dZ_{t}^{i} dZ_{t}^{j})=\rho_{l}^{2} \, dt,\,\,   \, \, i\ne j, \, \,    \nonumber\\[3mm]
&&\E(dW_{t}^{i} dW_{t}^{i})=\E(dZ_{t}^{j} dZ_{t}^{j})=dt,   \qquad  
 \E(dW_{t}^{i} dZ_{t}^{j})=0, \, \, \nonumber  \\[3mm] 
 &&\hskip6.5truecm t>0, \quad i,j=1,2,\ldots,N,
\end{eqnarray}
where $\E(\cdot)$ denotes the expected value of $\cdot$, and $\rho_{a}=\rho_{a,t}$,  $t>0$,  $\rho_{l}=\rho_{l,t}$,  $t>0$,  are  piecewise  constant   functions  of time   such that   $ |\rho_{a}| \le 1$,   $ |\rho_{l}| \le 1$, $t>0$. 
The stochastic differentials $dW_{t}^{i},$ $ t>0$, $ i=1,2,\ldots,N,$   can  be represented as follows:
\begin{equation}\label{rhoa}
dW_{t}^{i}=\rho_{a} d\widetilde{W}_{t}^{0}  +\sqrt{1-\rho_{a}^{2}}   \, d \widetilde{W}_{t}^{i},  \quad  t>0, \,  i=1,2,\ldots,N ,
\end{equation}
where $\widetilde{W}_{t}^{j}$, $  t\ge0$,   $j=0,1,\ldots,N$,  are independent standard Wiener processes such that $\widetilde{W}_{0}^{j}=0$,   $j=0,1,\ldots,N$,   and $d\widetilde{W}_{t}^{j}$, $  t>0$,   $j=0,1,\ldots,N$,  are their stochastic differentials.
The  term $ d \widetilde{W}_{t}^{0}$, $t>0$, is  called  common noise of the assets  equations (\ref{al1}).
Similarly   the stochastic  differentials  $dZ_{t}^{i},$ $ t>0$, $ i=1,2,\ldots,N,$  can  be represented as follows:
\begin{equation}\label{rhol}
d Z_{t}^{i}=\rho_{l} d \widetilde{Z}_{t}^{0}  +\sqrt{1-\rho_{l}^{2}}   \,  d \widetilde{Z}_{t}^{i},  \quad  t>0, \,  i=1,2,\ldots,N ,
\end{equation}
where $\widetilde{Z}_{t}^{j}$, $  t\ge0$,   $j=0,1,\ldots,N$,  are independent   standard Wiener processes such that  $\widetilde{Z}_{0}^{j}=0$,   $j=0,1,\ldots,N$,  and  $d\widetilde{Z}_{t}^{j}$, $  t>0$,   $j=0,1,\ldots,N$,  are their stochastic differentials. The  term  $ d \widetilde{Z}_{t}^{0}$, $t>0$, is called  common noise of the liabilities equations  (\ref{al2}).
Finally we assume that  $d \widetilde{W}_{t}^{i}$ and $ d \widetilde{Z}_{t}^{j}$ are independent,  $ t>0$, $ i,j=0,1,\ldots,N$.

Note  that   in  (\ref{rho})   the correlation coefficients    $\rho_{a}^{2}$, $\rho_{l}^{2}$ 
  between  the stochastic differentials of  the  assets equations (\ref{al1}) and of  the  liabilities equations      (\ref{al2}) are non negative.
These non negative correlation coefficients  generate  the so called  ``collective''  behaviour  of  the  banks  in presence of  a shock and  are
 translated in  the representation formulae  of the stochastic differentials (\ref{rhoa}),  (\ref{rhol}).
The  correlation  model   (\ref{rho})    can be  easily extended to more general situations.  
In this case   the representation formulae   (\ref{rhoa}),  (\ref{rhol}) must  be adapted to the circumstances. 
For simplicity  we omit these generalizations here.

  Note that   the diffusion coefficient $\sigma_{a}$  is the same  in  all the assets  equations  (\ref{al1})   and that  similar statements hold for the diffusion coefficient $\sigma_{l}$,    for the drift coefficients $\mu_{a}$, $\mu_{l}$ and for the correlation coefficients $\rho_{a}$, $\rho_{l}$.  
Moreover let us  assume that: $\tilde{a}_{0}^{i}=\tilde{a}_{0}$, $\tilde{l}_{0}^{i}=\tilde{l}_{0}$, $ i=1,2,\ldots,N$,  so that   we have:   $\tilde{a}_{0} >0$, $\tilde{l}_{0} >0$, $\tilde{a}_{0} - \tilde{l}_{0} >0$. 
With these   assumptions   all the banks of the model are  equal,  that  is  the bank population described by  the banking system model  (\ref{reserves}), (\ref{al1}), (\ref{al2}),  (\ref{alCIgeneral}),   (\ref{rho})  is homogeneous.  
Systems made of a homogeneous population of ``individuals''  are studied in statistical mechanics where the individuals are  usually atoms or molecules. 
In particular  extending the ideas developed in statistical mechanics to the study of banking system models we show that
the homogeneity of the bank population   implies    that,   when $N$ goes to infinity,  all the  banks     behave in the same way, that is  all the banks  behave  as  a kind of 
    ``mean bank'' and, using  the language  of statistical mechanics,  the    ``mean bank''  behaviour  is  defined by   the   ``mean field''  approximation   of the banking system model. 

In  an assets-liabilities dynamical  model  of   banking system  (like  model (\ref{reserves}), (\ref{al1}), (\ref{al2}),  (\ref{alCIgeneral}),   (\ref{rho}))   it is possible  to study    the propagation  of   certain types of shocks.  For example it is possible to model   shocks consisting   in  losses of  value of the    external assets  of the banks  caused  by a generalized fall  of the  market prices of the assets  and/or by a  generalized rise of the  expected defaults   (see, for example, \cite{Haldane},  \cite{May}).  
These shocks    reduce  the net worth of all  the banks at the same time  determining  an abrupt  increment    of   the probability  of   systemic risk  in a bounded time interval. 
In  model  (\ref{reserves}), (\ref{al1}), (\ref{al2}),  (\ref{alCIgeneral}),   (\ref{rho})        shocks  are 	 modeled  with  jumps  of  the volatility $\sigma_{a}$  in  the  assets equations (\ref{al1})     leaving  $\sigma_{l}$  constant in  the  liabilities   equations (\ref{al2})  or   viceversa
  with  jumps  of  $\sigma_{l}$      leaving  $\sigma_{a}$  constant.  
  For simplicity we do not consider jumps  of  $\sigma_{a}$ and  $\sigma_{l}$ occurring at the same time.
That is the shocks acting on the  assets and on  the  liabilities of the banks are  modeled  choosing the functions   $\sigma_{a}=\sigma_{a,t}$, $t>0$,   and      $\sigma_{l}=\sigma_{l,t}$, $t>0$.
Moreover in  model  (\ref{reserves}), (\ref{al1}), (\ref{al2}),  (\ref{alCIgeneral}),   (\ref{rho})  it is possible to study the 
 ``collective''   behaviour of  the  banks in presence  of a shock.  
In fact when a shock hits  the  banking system all  the banks  react in the same way and 
this ``collective''  behaviour  of  the  banks  is   modeled with  a positive correlation of the stochastic differentials on the right hand side of  
 the  assets equations (\ref{al1}) and/or  of the  liabilities equations      (\ref{al2}).
 That is the  ``collective''   behaviour of the banks in reaction  to a shock   is modeled  with a jump of     the functions $\rho_{a}=\rho_{a,t}$, $t>0$,    and/or     $\rho_{l}=\rho_{l,t}$, $t>0$.

Let us  adapt  to    model  (\ref{reserves}), (\ref{al1}), (\ref{al2}),  (\ref{alCIgeneral}),   (\ref{rho})  the mechanisms  used  in the models presented in  \cite{Fouque1}, \cite{nostro1}  to describe     the cooperation among  banks and let us  introduce the terms used to describe 
the intervention of the   monetary authority in the   banking system  dynamics. 
%
To do this we  define      the  new   variables  $G_{t}^{i}$,  $H_{t}^{i}$, $t\ge0$, $i=1,2,\ldots,N$, as follows:
\begin{equation} \label{alcoupled1}
G_{t}^{i}=\ln(a_{t}^{i}), \quad H_{t}^{i}=\ln(l_{t}^{i}),  \quad  t\ge0, \quad i=1,2,\ldots,N, 
\end{equation}
where  $\ln(\cdot)$ is the logarithm of $\cdot$. 
First of all note that   the variables 
 $G_{t}^{i}=\ln(a_{t}^{i})$,  $H_{t}^{i}=\ln(l_{t}^{i})$, $t\ge0$, $i=1,2,\ldots,N$, are well-defined. 
 In fact,  at time $t=0$,  for $ i=1,2,\ldots,N$,  we have  $\tilde{a}_{0}^{i}=\tilde{a}_{0}>0$, 
 $\tilde{l}_{0}^{i}=\tilde{l}_{0} >0$,  $\tilde{a}_{0} -\tilde{l}_{0} >0$, with probability one,  and   therefore   equations   (\ref{al1}), (\ref{al2}) imply  that    $a_{t}^{i}>0$, $l_{t}^{i}>0$, with probability one, $t>0$.

The quantities   $G_{t}^{i}=\ln(a_{t}^{i})$,  $H_{t}^{i}=\ln(l_{t}^{i})$,    are, respectively,  the
    log-assets   and   the  log-liabilities  of the  $i$-th bank  at time  $t\ge0$, $i=1,2,\ldots,N$.   
 
  Using  It\^{o}'s  Lemma  and equations  (\ref{al1}), (\ref{al2})  it is easy to see  that the   stochastic processes   $G_{t}^{i}$,  $H_{t}^{i}$, $t\ge0$, $i=1,2,\ldots,N$, satisfy the  following  equations:
\begin{eqnarray}
&& d G_{t}^{i}= \left(\mu_{a}  -\frac12\sigma_{a}^{2}\right) dt +  \sigma_{a}  dW_{t}^{i} , \quad    t>0, \,  i=1,2,\ldots,N ,  \label{logal1}\\
&& d H_{t}^{i}= \left(\mu_{l}  -\frac12\sigma_{l}^{2}\right) dt +  \sigma_{l}  dZ_{t}^{i},\quad    t>0, \,  i=1,2,\ldots,N ,  \label{logal2}
\end{eqnarray}
%
%
%
%
%
%
%
%
%
%
and  the initial conditions:  
\begin{eqnarray}\label{logalCIgeneral}
G_{0}^{i}=\ln(\tilde{a}_{0}^{i})=\ln(\tilde{a}_{0}), \quad H_{0}^{i}=\ln(\tilde{l}_{0}^{i})=\ln(\tilde{l}_{0}), \quad i=1,2,\ldots,N.
\end{eqnarray}
%
%
Let us define      the stochastic processes: 
\begin{eqnarray}
&&A_{t}^{i}= G_{t}^{i}-\left(\mu_{a}  t   -\frac12 \int_{0}^{t}\sigma_{a,\tau}^{2} d \tau\right), \quad  t \ge 0, \,  i=1,2,\ldots,N ,\\
&&L_{t}^{i}=H_{t}^{i}-\left(\mu_{l}  t   -\frac12 \int_{0}^{t}\sigma_{l,\tau}^{2} d \tau\right),  \quad t \ge 0, \,  i=1,2,\ldots,N ,
\end{eqnarray}
 from (\ref{logal1}), (\ref{logal2}), (\ref{logalCIgeneral}) it  is easy to see  that $A_{t}^{i}$, $L_{t}^{i}$, $t\ge0$, $  i=1,2,\ldots,N,$ satisfy the  following  equations:
\begin{eqnarray}
%
&& d A_{t}^{i}=  \sigma_{a}  dW_{t}^{i} ,\quad    t>0, \,  i=1,2,\ldots,N ,  \label{ALnewInd1}\\
&& d L_{t}^{i}=  \sigma_{l}  dZ_{t}^{i} ,\quad    t>0, \,  i=1,2,\ldots,N ,  \label{ALnewInd2}
\end{eqnarray}
and the   initial conditions:  
\begin{eqnarray}\label{ALnewIndCI}
A_{0}^{i}=\ln(\tilde{a}_{0}^{i})=\ln(\tilde{a}_{0}), \quad L_{0}^{i}=\ln(\tilde{l}_{0}^{i})=\ln(\tilde{l}_{0}), \quad i=1,2,\ldots,N.
\end{eqnarray}

Let  $\psi_{t}$,  $ t\ge0$, be   a continuous piecewise differentiable function,  the  notation  
 $\displaystyle d\psi_{t}=\frac{d\psi_{t}}{dt} \,dt=(\psi_{t})' \,dt$, $ t>0$,  denotes  the  ``piecewise differential'' of  $\psi_{t}$,  $ t\ge0$.

Using   the ideas developed  in   \cite{Fouque1}, \cite{nostro1}   we  modify the    equations     (\ref{ALnewInd1}), (\ref{ALnewInd2}) and we introduce  the terms used to  implement the  cooperation mechanism  among  banks and   the terms used to model the intervention of the monetary authority in the banking system dynamics. This is done     adding  to     (\ref{ALnewInd1}), (\ref{ALnewInd2})  some drift terms. 
That is, given the continuous piecewise differentiable  functions   $\varphi_{t} >0$,   $\phi_{t} >0$,   $ t \ge 0$,  such that $\varphi_{t}- \phi_{t} >0$,   $ t \ge 0$,  we replace   equations  (\ref{ALnewInd1}), (\ref{ALnewInd2}), respectively,   with the    equations:
\begin{eqnarray}
&& d A_{t}^{i}=  \frac{\alpha_{t}}{N}\sum_{j=1}^{N} \left( A_{t}^{j} -A_{t}^{i}\right) dt   +  d \tilde{\varphi}_{t} + \sigma_{a}  dW_{t}^{i} , \quad t>0, \,  i=1,2,\ldots,N ,  \label{ALnew1}\\
&&  d L_{t}^{i}=  \frac{\gamma_{t}}{N}\sum_{j=1}^{N} \left( L_{t}^{j} -L_{t}^{i}\right) dt  + d \tilde{\phi}_{t}  + \sigma_{l}  dW_{t}^{i} ,   \quad t>0, \,  i=1,2,\ldots,N ,  \label{ALnew2}
\end{eqnarray}
where  the functions $\tilde{\varphi}_{t}$, $\tilde{\phi}_{t}$, $t\ge0$, are given by:
\begin{eqnarray}
&&
\tilde{\varphi}_{t}=\ln(\varphi_{t})- \mu_{a}  t   +\frac12 \int_{0}^{t}\sigma_{a, \tau}^{2} d \tau,   \quad t\ge0,  \label{funvarphi}\\
&& \tilde{\phi}_{t}=\ln(\phi_{t})- \mu_{l}  t   +\frac12 \int_{0}^{t}\sigma_{l, \tau}^{2} d \tau , \quad  t\ge0.\label{funphi}
\end{eqnarray}

The equations  (\ref{ALnew1}), (\ref{ALnew2})  are completed with the initial conditions   (\ref{ALnewIndCI}) and with the assumptions on the correlation  coefficients    (\ref{rho}).  For later convenience  from now on  we assume  that: $\tilde{a}_{0}=\varphi_{0}$,   $\tilde{l}_{0}=\phi_{0}$.

The  equations  (\ref{ALnew1}), (\ref{ALnew2})  are,  respectively, the equations that describe  the ``assets''  and     the ``liabilities''  of the banks.  
For  simplicity    the variables $A_{t}^{i}$,  $L_{t}^{i}$, $t\ge0$,  are called   respectively   ``assets'' and ``liabilities''
  of the $i$-th bank, $i=1,2,\ldots,N$,   instead of  being called  centered log-assets  and  centered log-liabilities as it should be more appropriate.

  The functions  $\varphi_{t}$,   $\phi_{t}$,   will be interpreted,  respectively,  as  assets and  liabilities of the ``ideal bank'' at time $t$,  $ t \ge 0$.  The fact that the ``ideal bank''  is  solvent   corresponds to the assumption   that $\varphi_{t}-\phi_{t} >0$,  $ t \ge 0$.
Recall  that the functions      $\tilde{\varphi}_{t}$, $\tilde{\phi}_{t}$, $t \ge 0$,  of equations    (\ref{ALnew1}), (\ref{ALnew2}) are related to $\varphi_{t}$,   $\phi_{t}$,  
  $ t \ge 0$,  through  (\ref{funvarphi}), (\ref{funphi}).
 The functions  $\alpha_{t} \ge 0$, $\gamma_{t} \ge 0$,  $t>0$,  of (\ref{ALnew1}), (\ref{ALnew2})  regulate  
  the cooperation mechanism   among  banks and their  choice   corresponds to the choice of the rules  of the cooperation mechanism   among  banks.
 Later this choice  will be attributed to the monetary authority
 and will be used  to govern   the systemic risk  in a bounded time interval  of   the banking system model.
 The initial value problem  (\ref{ALnew1}), (\ref{ALnew2}), (\ref{ALnewIndCI}) is completed with the assumptions   (\ref{rho}). 

For  $i=1,2,\ldots,N$     the  cooperation  of  the $i$-th  bank  with the other banks   is described  by the  drift terms   
$\displaystyle  \frac{\alpha_{t}}{N}\sum_{j=1}^{N} \left( A_{t}^{j} -A_{t}^{i}\right) dt$, $t>0$,  and
$\displaystyle  \frac{\gamma_{t}}{N}\sum_{j=1}^{N} \left( L_{t}^{j} -L_{t}^{i}\right) dt$, $t>0$,   respectively,   of the  $i$-th equation    (\ref{ALnew1}) and of  the  $i$-th equation  (\ref{ALnew2}).
In fact   the term  $\displaystyle  \frac{\alpha_{t}}{N}\sum_{j=1}^{N} \left( A_{t}^{j} -A_{t}^{i}\right) dt$, in  the  $i$-th equation    (\ref{ALnew1})  implies   that  for  $t>0$  and  $ j=1,2,\ldots,N$,  if at time $t$   bank $j$ has more ``assets'' than  bank $i$ (i.e. if  $ A_{t}^{j} >A_{t}^{i}$)  assets  flow from bank $j$ to  bank $i$, and this  flow is  proportional to the difference  $A_{t}^{j} -A_{t}^{i}$  at the  rate  $\displaystyle \frac{\alpha_{t}}{N}$,   the opposite happens if   bank $i$ has more ``assets''  than  bank $j$
 (i.e. if  $ A_{t}^{j} <A_{t}^{i}$),  $j=1,2,\ldots,N$.
 For $i=1,2,\ldots,N$, the   term   $\displaystyle  \frac{\gamma_{t}}{N}\sum_{j=1}^{N} \left( L_{t}^{j} -L_{t}^{i}\right) dt$,  $t>0$,  in  the  $i$-th equation    (\ref{ALnew2}),     
is relative to the ``liabilities''  and is analogous of the term
 $\displaystyle  \frac{\alpha_{t}}{N}\sum_{j=1}^{N} \left( A_{t}^{j} -A_{t}^{i}\right) dt$,  $t>0$,   of the ``assets'' of  the  $i$-th  equation    (\ref{ALnew1}); 
  this term   has  the same effect  on the liabilities   than
 the  effect   that  the term 
 $\displaystyle  \frac{\alpha_{t}}{N}\sum_{j=1}^{N} \left( A_{t}^{j} -A_{t}^{i}\right) dt$,  $t>0$,  has on  the assets.  
 
 Note that the division by $N$ in  the rates $\displaystyle \frac{\alpha_{t}}{N}$, $ \displaystyle\frac{\gamma_{t}}{N}$, $t>0$,    of the drift terms  of   equations  (\ref{ALnew1}), (\ref{ALnew2})   is a normalization factor taken from  the technical  literature (see, for example,     \cite{Fouque2},  \cite {Papan1}, \cite{nostro1})
  that  plays no role in this paper.

  The     cooperation    mechanism   added in    (\ref{ALnew1}), (\ref{ALnew2})    is  a simple implementation of the idea that  ``who has more (assets, liabilities) gives to those who have less (assets, liabilities)'',  in this sense it is a  cooperation mechanism  among  banks (see  \cite{nostro1}).

 The drift  terms $ d \tilde{\varphi}_{t}$, $d \tilde{\phi}_{t}$,   $t>0$,     of    equations  (\ref{ALnew1}), (\ref{ALnew2})   describe   the intervention of the monetary authority in the banking system dynamics.
 In fact  the  term  $ d \tilde{\varphi}_{t}$, $t>0$,   of  the equations  (\ref{ALnew1}) is responsible for the fact  that the drift terms   $\displaystyle  \frac{\alpha_{t}}{N}\sum_{j=1}^{N} \left( A_{t}^{j} -A_{t}^{i}\right) dt$, $t>0$, $i=1,2,\ldots,N, $ stabilize  the trajectories   of  $A_{t}^{i}$, $t>0$,  $i=1,2,\ldots,N, $ around  the function 
 $  \tilde{\varphi}_{t}$, $t>0$,  and, as a consequence, stabilize  the trajectories   of $a_{t}^{i}$, $t>0$,  $i=1,2,\ldots,N, $ around  the function  $\varphi_{t}$, $t>0$.
Analogously  the term  $ d \tilde{\phi}_{t}$, $t>0$,  of the equations    (\ref{ALnew2})  is responsible for the fact  that the drift terms   $\displaystyle  \frac{\gamma_{t}}{N}\sum_{j=1}^{N} \left( L_{t}^{j} -L_{t}^{i}\right) dt$, $t>0$, $i=1,2,\ldots,N, $ stabilize  the trajectories   of  $L_{t}^{i}$, $t>0$,  $i=1,2,\ldots,N, $ around  the function $  \tilde{\phi}_{t}$, $t>0$,  and, as a consequence, stabilize  the trajectories   of $l_{t}^{i}$, $t>0$,  $i=1,2,\ldots,N, $ around the function  $\phi_{t}$, $t>0$.
That is      when $\alpha_{t}>0$, $\gamma_{t}>0$, $t>0$,   the drift terms  introduced  
in  equations    (\ref{ALnew1}), (\ref{ALnew2}) 
to model the cooperation  mechanism among   banks and    the intervention of the monetary authority in the  
 banking system dynamics  expressed by the terms    $ d \tilde{\varphi}_{t}$, $d \tilde{\phi}_{t}$,   $t>0$,  generate a ``swarming'' effect of the trajectories of the assets $a_{t}^{i}$, $t>0$,  $i=1,2,\ldots,N, $ and of the liabilities $l_{t}^{i}$, $t>0$,  $i=1,2,\ldots,N, $  around, respectively,  $  \varphi_{t}$, $  \phi_{t}$, $t>0$, that is  around, respectively, the assets and the liabilities of the ``ideal bank''. 
This implies that the trajectories of the capital reserves of the $i$-th bank swarms around the  capital reserves
 of the ``ideal bank'' $  \xi_{t}=\varphi_{t}- \phi_{t}$, $t>0$, $i=1,2,\ldots,N$.
  This  swarming effect  is a key ingredient  of the systemic risk governance  discussed later.

Let us rewrite equations  (\ref{ALnew1}), (\ref{ALnew2}), (\ref{ALnewIndCI}) using as dependent variables the stochastic processes   $G_{t}^{i}$, $H_{t}^{i},$ $ t\ge0, $   
 $i=1,2,\ldots,N$.  We have:
%
%
%
\begin{eqnarray}
&& d G_{t}^{i}=  \frac{\alpha_{t}}{N}\sum_{j=1}^{N} \left( G_{t}^{j} -G_{t}^{i}\right) dt  + d \ln(\varphi_{t}) + \sigma_{a}  dW_{t}^{i} , \quad t\ge0, \,  i=1,2,\ldots,N ,  \label{alcoupled2}\\
&& d H_{t}^{i}=  \frac{\gamma_{t}}{N}\sum_{j=1}^{N} \left( H_{t}^{j} -H_{t}^{i}\right) dt  +d \ln(\phi_{t}) + \sigma_{l}  dZ_{t}^{i} ,\quad t\ge0, \,  i=1,2,\ldots,N ,  \label{alcoupled3}
\end{eqnarray}
with the  initial conditions:
\begin{eqnarray}\label{logalCI}
G_{0}^{i}=\ln(\tilde{a}_{0}), \quad H_{0}^{i}=\ln(\tilde{l}_{0}), \quad i=1,2,\ldots,N,
\end{eqnarray}
where     $\tilde{a}_{0}=\varphi_{0}$,   $\tilde{l}_{0}=\phi_{0}$. To the equations  (\ref{reserves}), (\ref{alcoupled1}), (\ref{alcoupled2}), (\ref{alcoupled3}), (\ref{logalCI})  is added  the assumption (\ref{rho}), this completes the banking system model.

For simplicity we use     the same  symbols   to denote  the variables  of    model  (\ref{reserves}), (\ref{al1}), (\ref{al2}),  (\ref{alCIgeneral}),   (\ref{rho}) and   those of    model      (\ref{reserves}), (\ref{alcoupled1}), (\ref{alcoupled2}), (\ref{alcoupled3}), (\ref{logalCI}),   (\ref{rho}).   When necessary to  avoid ambiguity  we specify  the banking system  model considered.

Note that when  $\alpha_{t} =0$, $\gamma_{t} =0$,  $t>0$, and the functions  $\varphi_{t}$, $\phi_{t}$, $t \ge 0$,  are constants,  there is no cooperation  among banks and no  intervention of the  monetary authority in the banking system dynamics.  
In this case  model (\ref{reserves}), (\ref{alcoupled1}), (\ref{alcoupled2}), (\ref{alcoupled3}), (\ref{logalCI}),   (\ref{rho}) reduces to model (\ref{reserves}), (\ref{al1}), (\ref{al2}),  (\ref{alCIgeneral}),   (\ref{rho}).

%


\section{Systemic risk in a bounded time interval }\label{sec3}
%
%
Given  the    banking system model   (\ref{reserves}), (\ref{al1}), (\ref{al2}),  (\ref{alCIgeneral}),   (\ref{rho}),   or  the    banking system model   
 (\ref{reserves}), (\ref{alcoupled1}), (\ref{alcoupled2}), (\ref{alcoupled3}), (\ref{logalCI}),   (\ref{rho}),   
 we define  the  events: i) default of a bank in a bounded time interval, ii)  systemic risk  in a bounded time interval 
 and  we introduce a probability distribution   called   loss distribution  of the banks defaulted  in a bounded time interval.

%
%
Given $ 0\le \tau_{1}<\tau_{2} < + \infty$  and  the default level   $D\ge 0$     we define the  event $F^{i}_{[\tau_{1},\tau_{2}]}$,     ``default   of the $i$-th bank in  the time interval $[\tau_{1},\tau_{2}]$'', as follows:
\begin{eqnarray}  \label{failure1}
&&  F^{i}_{[\tau_{1},\tau_{2}]}  =\left\{  \min_{\tau_{1}\le t\le \tau_{2}}  c_{t}^{i} < D  \right\} , \quad i=1,2,\ldots,N.
\end{eqnarray}
That is for $i=1,2,\ldots,N$   the $i$-th bank    defaults  in the  time interval $[\tau_{1},\tau_{2}]$   if  in that time interval its capital reserves    $c_{t}^{i}$, $t\ge0$,   go below   the default  level $D$.   Recall that in this paper   we have chosen     $D=0$ and that the inequality 
$\displaystyle  \min_{\tau_{1}\le t\le \tau_{2}}  c_{t}^{i} < D  $  is considered  on each  path of the stochastic process 
$ c_{t}^{i}$, $\tau_{1}\le t\le \tau_{2}$,  $i=1,2,\ldots,N$. 
The failed  banks  are   removed from   the   banking system  model,  this means that the  number of   banks present in   the    model 
may depend from the path of the banking system  model considered and may not be constant during the time evolution.

Let  $int\left[\cdot\right]$ be  the integer part of the real number $\cdot$, and $M$ be a positive integer such that  $ int\left[\frac{N}{2}\right] \le M \le N$. 
  The systemic risk  (or  systemic event)   of type $M$ in  the time interval $[\tau_{1},\tau_{2}]$,   $SR_{[\tau_{1},\tau_{2}]}$, is  the event  defined as follows:
\begin{equation}  \label{SR}
SR_{[\tau_{1},\tau_{2}]}^{M}=\left\{   \mbox{at least} \,  M  \mbox{  banks fail in  the time interval } \, [\tau_{1},\tau_{2}]\right\} .   
\end{equation}
In this paper   we   choose $M = int\left[\frac{N}{2}\right]+1$ and we write $SR_{[\tau_{1},\tau_{2}]}$ to mean $SR_{[\tau_{1},\tau_{2}]}^{M}$
when  $M=int\left[\frac{N}{2}\right]+1$.

Let $\pr(\cdot)$ be the probability of the event  $\cdot$. 
Given the banking system model 
(\ref{reserves}), (\ref{al1}), (\ref{al2}),  (\ref{alCIgeneral}),   (\ref{rho})
or
 (\ref{reserves}), (\ref{alcoupled1}), (\ref{alcoupled2}), (\ref{alcoupled3}), (\ref{logalCI}),   (\ref{rho})
  to the events  $F^{i}_{[\tau_{1},\tau_{2}]}$, $i=1,2,\ldots,N$, and $SR_{[\tau_{1},\tau_{2}]}$ defined in 
  (\ref{failure1}), (\ref{SR})   it is associated a probability that  is  evaluated    using statistical simulation. 
In fact     the probability  $\pr(F^{i}_{[\tau_{1},\tau_{2}]})$   of   the event    $F^{i}_{[\tau_{1},\tau_{2}]}$,   $i=1,2,\ldots,N, $  and  the probability
 $\pr(SR_{[\tau_{1},\tau_{2}]})$ of the event $SR_{[\tau_{1},\tau_{2}]}$    is  approximated    with the  corresponding frequencies  computed   on  a set of numerically simulated trajectories of the banking system model considered.
 Note that due to the homogeneity of the bank population  $\pr(F^{i}_{[\tau_{1},\tau_{2}]})$  does not depend on $i$ when $i=1,2,\ldots,N$.

The loss distribution   of the banks   defaulted  in the  bounded  time interval  $[\tau_{1},\tau_{2}]$  is the probability distribution  of the random variable: number of bank defaults   in the  time interval $[\tau_{1},\tau_{2}]$. Given a  banking system model the loss distribution of the banks   defaulted   in the time interval  $[\tau_{1},\tau_{2}]$  can be   approximated      using  statistical simulation  computing   the    distribution  of the frequencies of the appropriate    events in a set of numerically simulated trajectories of the banking system model considered.

\begin{figure}
 \centerline{\includegraphics[height=6cm]{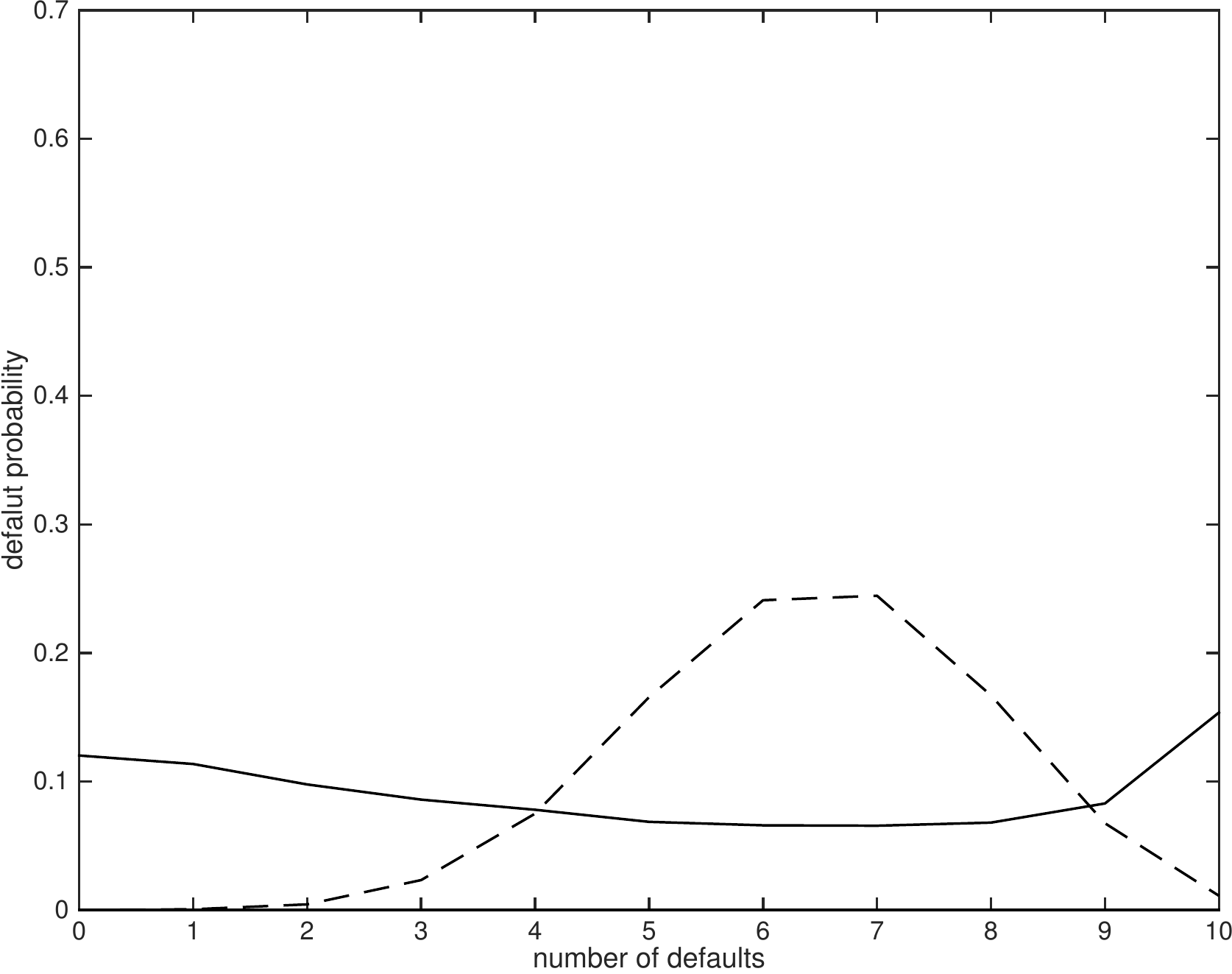}}
   \caption{ \small Loss distribution in  $[0,T]$,  $T=1$,  of system  (\ref{reserves}), (\ref{alcoupled1}), (\ref{alcoupled2}), (\ref{alcoupled3}), (\ref{logalCI}),   (\ref{rho}) when  $N=10$, $ \mu_{a}=0.1$,  $ \mu_{l}=0.1$,  $ \sigma_{a}=\sigma_{1,t}$, $ \sigma_{l}=0.6$,   $\rho_{a}=0$, $ \rho_{l}=0$,  $\varphi_{t}=0.1$,  $\phi_{t}  =0.06$, $\alpha_{t}=10,$  $\gamma_{t}=10$, $t\in[0,T]$,  $T=1$ (solid line) 
   and  loss distribution in  $[0,T]$,  $T=1$,  of   system  (\ref{reserves}), (\ref{al1}), (\ref{al2}),  (\ref{alCIgeneral})  when $N=10$,  $ \mu_{a}=0.1$,  $ \mu_{l}=0.1$,  $ \sigma_{a}=\sigma_{1,t}$,  $ \sigma_{l}=0.6$,   $\rho_{a}=0$, $ \rho_{l}=0$, $t\in[0,T]$,  $T=1$ (dashed  line).}
 \label{fig1}
\vskip1truecm
%
 \centerline{\includegraphics[height=6cm]{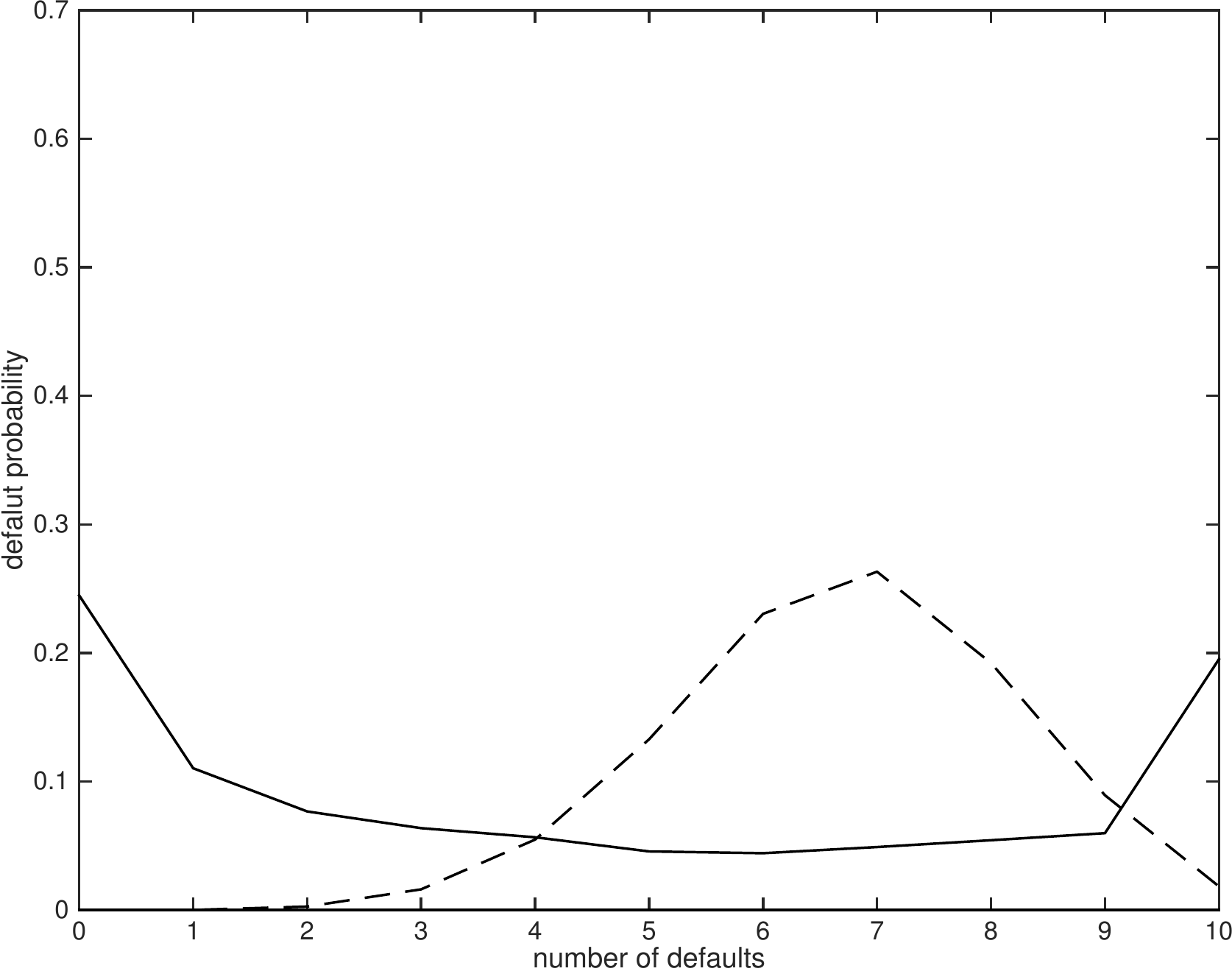}}
  \caption{  \small Loss distribution in  $[0,T]$,  $T=1$,  of system  (\ref{reserves}), (\ref{alcoupled1}), (\ref{alcoupled2}), (\ref{alcoupled3}), (\ref{logalCI}),   (\ref{rho}) when  $N=10$, $ \mu_{a}=0.1$,  $ \mu_{l}=0.1$,  $ \sigma_{a}=\sigma_{2,t}$,  $ \sigma_{l}=0.6$,   $\rho_{a}=0$, $ \rho_{l}=0$,  $\varphi_{t}=0.1$,  $\phi_{t}  =0.06$, $\alpha_{t}=20,$  $\gamma_{t}=20$, $t\in[0,T]$,  $T=1$ (solid line) 
   and  loss distribution in  $[0,T]$,  $T=1$,  of   system  (\ref{reserves}), (\ref{al1}), (\ref{al2}),  (\ref{alCIgeneral})  when $N=10$,  $ \mu_{a}=0.1$,  $ \mu_{l}=0.1$,  $ \sigma_{a}=\sigma_{2,t}$,  $ \sigma_{l}=0.6$,   $\rho_{a}=0$, $ \rho_{l}=0$, $t\in[0,T]$,  $T=1$ (dashed  line).}
 \label{fig2}
  \end{figure}

\begin{figure}
 \centerline{\includegraphics[height=6cm]{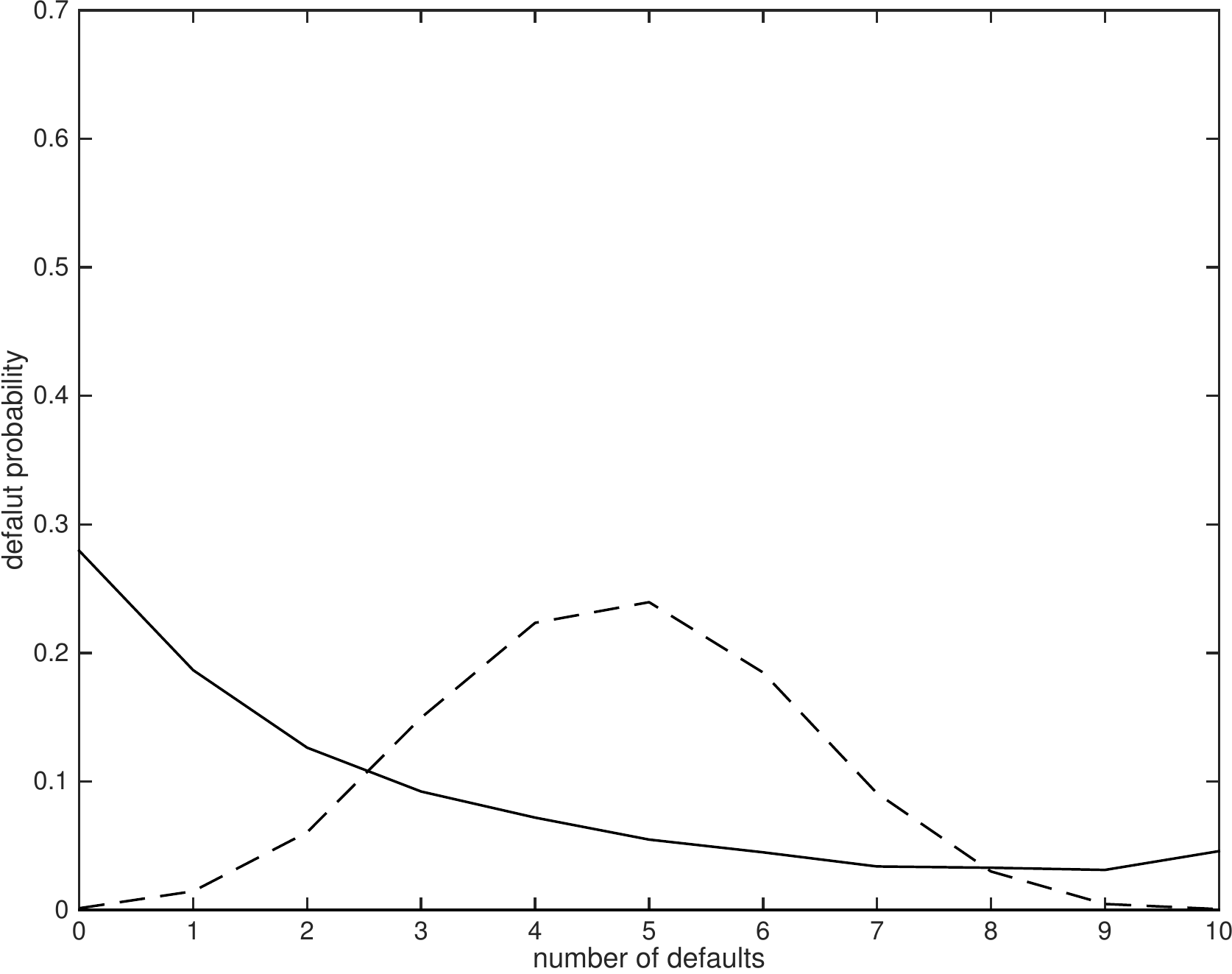}}
   \caption{ \small Loss distribution in  $[0,T]$,  $T=1$,  of system  (\ref{reserves}), (\ref{alcoupled1}), (\ref{alcoupled2}), (\ref{alcoupled3}), (\ref{logalCI}),   (\ref{rho}) when  $N=10$, $ \mu_{a}=0.1$,  $ \mu_{l}=0.1$,  $ \sigma_{a}=\sigma_{3,t}$,  $ \sigma_{l}=0.6$,   $\rho_{a}=0$, $ \rho_{l}=0$,  $\varphi_{t}=0.1$,  $\phi_{t}  =0.06$, $\alpha_{t}=10,$  $\gamma_{t}=10$, $t\in[0,T]$,  $T=1$  (solid line) 
   and  loss distribution in  $[0,T]$,  $T=1$,  of   system  (\ref{reserves}), (\ref{al1}), (\ref{al2}),  (\ref{alCIgeneral})  when $N=10$,  $ \mu_{a}=0.1$,  $ \mu_{l}=0.1$,  $ \sigma_{a}=\sigma_{3,t}$,  $ \sigma_{l}=0.6$,   $\rho_{a}=0$, $ \rho_{l}=0$, $t\in[0,T]$,  $T=1$  (dashed  line).}
 \label{fig3}
  \end{figure}

\begin{figure}
   \centerline{\includegraphics[height=6cm]{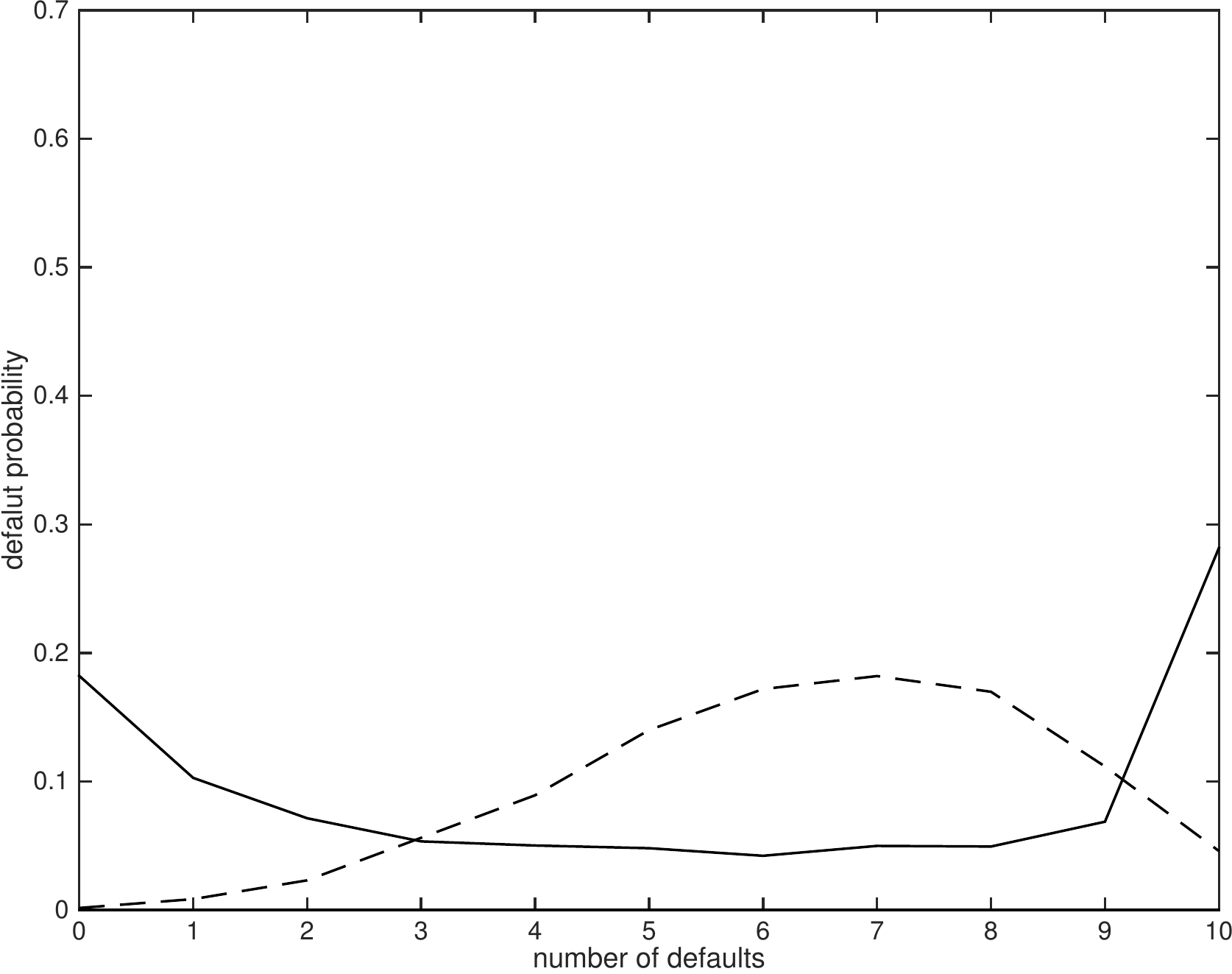} }
   \caption{\small  Loss distribution in  $[0,T]$,  $T=1$,  of system  (\ref{reserves}), (\ref{alcoupled1}), (\ref{alcoupled2}), (\ref{alcoupled3}), (\ref{logalCI}),   (\ref{rho}) when  $N=10$, $ \mu_{a}=0.1$,  $ \mu_{l}=0.1$,  $ \sigma_{a}=\sigma_{1,t}$,  $ \sigma_{l}=0.6$,   $\rho_{a}=0.5$, $ \rho_{l}=0$,  $\varphi_{t}=0.1$,  $\phi_{t}  =0.06$, $\alpha_{t}=10,$  $\gamma_{t}=10$, $t\in[0,T]$,  $T=1$ (solid line) 
   and  loss distribution in  $[0,T]$,  $T=1$,  of   system  (\ref{reserves}), (\ref{al1}), (\ref{al2}),  (\ref{alCIgeneral})  when $N=10$,  $ \mu_{a}=0.1$,  $ \mu_{l}=0.1$,  $ \sigma_{a}=\sigma_{1,t}$,  $ \sigma_{l}=0.6$,   $\rho_{a}=0.5$, $ \rho_{l}=0$, $t\in[0,T]$,  $T=1$ (dashed  line).}
 \label{fig4}
 \end{figure}

\begin{figure}
   \centerline{\includegraphics[height=6cm]{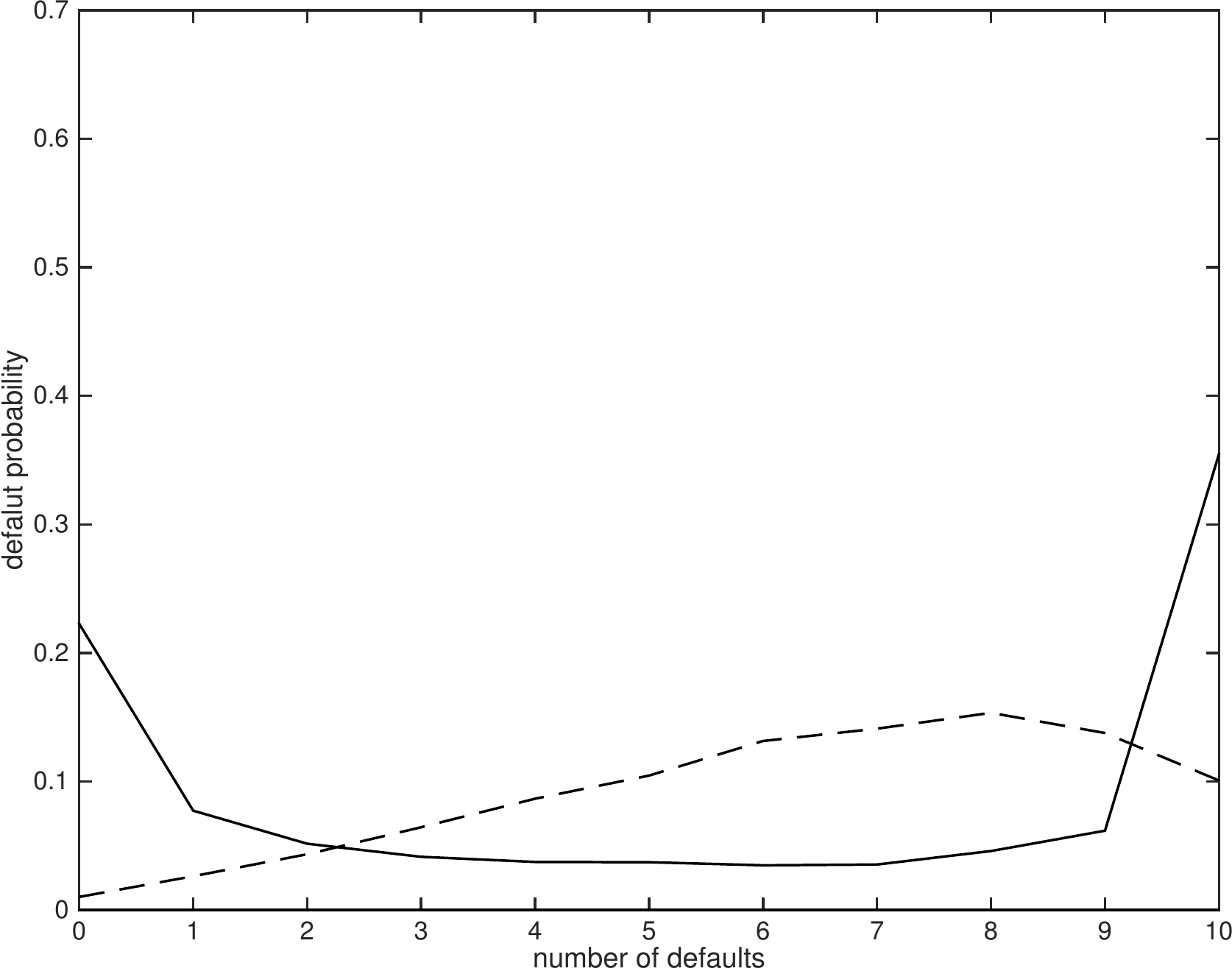} }
    \caption{\small  Loss distribution in  $[0,T]$,  $T=1$,  of system  (\ref{reserves}), (\ref{alcoupled1}), (\ref{alcoupled2}), (\ref{alcoupled3}), (\ref{logalCI}),   (\ref{rho}) when  $N=10$, $ \mu_{a}=0.1$,  $ \mu_{l}=0.1$,  $ \sigma_{a}=\sigma_{1,t}$,  $ \sigma_{l}=0.6$,   $\rho_{a}=\sqrt{0.5}$, $ \rho_{l}=0$,  $\varphi_{t}=0.1$,  $\phi_{t}  =0.06$, $\alpha_{t}=10,$  $\gamma_{t}=10$, $t\in[0,T]$,  $T=1$ (solid line) 
   and  loss distribution in  $[0,T]$,  $T=1$,  of   system  (\ref{reserves}), (\ref{al1}), (\ref{al2}),  (\ref{alCIgeneral})  when $N=10$,  $ \mu_{a}=0.1$,  $ \mu_{l}=0.1$,  $ \sigma_{a}=\sigma_{1,t}$,  $ \sigma_{l}=0.6$,   $\rho_{a}=\sqrt{0.5}$, $ \rho_{l}=0$, $t\in[0,T]$,  $T=1$ (dashed  line).}
 \label{fig5}
  \end{figure}

Let us  study   the loss distribution of the banks   defaulted   in the  time interval  $[0,T]$,  $T=1$,  in the banking system model (\ref{reserves}), (\ref{al1}), (\ref{al2}),  (\ref{alCIgeneral}),   (\ref{rho})
  and  in  the banking system model 
 (\ref{reserves}), (\ref{alcoupled1}), (\ref{alcoupled2}), (\ref{alcoupled3}), (\ref{logalCI}),   (\ref{rho}). 
 In both models we choose  $N=10$, and we  evaluate the   loss distribution  of the banks   defaulted   in  $[0,T]$,  $T=1$,    using statistical simulation    starting from  $10^{4}$ numerically simulated trajectories of the  models  considered. 
Let us define  the    functions:
\begin{equation}  \label{sigma1t}
\sigma_{1,t}=0.8,  \quad  t \in [0,1],
\end{equation}
\begin{equation}  \label{sigma2t}
\sigma_{2,t}=
\left\{%
\begin{array}{ll}
  \displaystyle 0.2 ,  & t\in [0,0.2]  , \\
  \displaystyle  1 ,  & t\in (0.2,1],
\end{array}%
\right.
\end{equation}
\begin{equation}  \label{sigma3t}
\sigma_{3,t}=
\left\{%
\begin{array}{ll}
  \displaystyle 0.2 ,  & t\in [0,0.2]  , \\
  \displaystyle  0.8 ,  & t\in (0.2, 0.5],\\ 
   \displaystyle  0.2 ,  & t\in (0.5,1] .
\end{array}%
\right.
\end{equation}
In Figures   \ref{fig1}-\ref{fig5} the dashed  line  shows the loss distribution of the banks   defaulted  in the time interval  $[0,T]$,  $T=1$, of  model  (\ref{reserves}), (\ref{al1}), (\ref{al2}),  (\ref{alCIgeneral}),   (\ref{rho}),  while the solid line shows the loss distribution  of the banks   defaulted   in  the time interval  $[0,T]$,  $T=1$, of  model  (\ref{reserves}), (\ref{alcoupled1}), (\ref{alcoupled2}), (\ref{alcoupled3}), (\ref{logalCI}),   (\ref{rho}) when
in  Figures     \ref{fig1}-\ref{fig5} we have:
 $N=10$, $\varphi_{t}=0.1$,  $\phi_{t}  =0.06$,   $ \sigma_{l}=\sigma_{l,t}=0.6$,  $ \rho_{l}= \rho_{l,t}=0$,  $t\in[0,T]$,  $T=1$, $ \mu_{a}=0.1$,  $ \mu_{l}=0.1$; moreover in  Figure    \ref{fig1} we have:
$ \sigma_{a}=\sigma_{1,t}$,   $\rho_{a}=\rho_{a,t}=0$,  $\alpha_{t}=10,$  $\gamma_{t}=10$, $t\in[0,T]$,  $T=1$, 
 in  Figure    \ref{fig2} we have: 
$ \sigma_{a}=\sigma_{2,t}$,   $\rho_{a}=\rho_{a,t}=0$,  $\alpha_{t}=20,$  $\gamma_{t}=20$,  $t\in[0,T]$,  $T=1$,
 in  Figure    \ref{fig3} we have: 
$ \sigma_{a}=\sigma_{3,t}$,   $\rho_{a}=\rho_{a,t}=0$,  $\alpha_{t}=10,$  $\gamma_{t}=10$, $t\in[0,T]$,  $T=1$, 
  in  Figure    \ref{fig4} we have: 
$ \sigma_{a}=\sigma_{1,t}$,   $\rho_{a}=\rho_{a,t}=0.5$,  $\alpha_{t}=10,$  $\gamma_{t}=10$, $t\in[0,T]$,  $T=1$,  and finally 
 in  Figure    \ref{fig5} we have: 
$ \sigma_{a}=\sigma_{1,t}$,   $\rho_{a}=\rho_{a,t}=\sqrt{0.5}$,  $\alpha_{t}=10,$  $\gamma_{t}=10$, $t\in[0,T]$,  $T=1$.

Note that  the results shown  in Figures \ref{fig1}-\ref{fig5}  are obtained  when  the functions   $\varphi_{t}=0.1$,  $\phi_{t}=0.06$, $t\in[0,T]$,  $T=1$,   are  constants.  With this choice   there is no  intervention of the monetary authority    in the  banking system dynamics  for  $t\in(0,T]$ (in fact  $d \ln(\varphi_{t}) =d \ln(\phi_{t}) =0$, $t\in(0,T]$,   in (\ref{alcoupled2}), (\ref{alcoupled3}))
and only the  cooperation mechanism among banks is active when $t\in[0,T]$.
Note that  in Figures \ref{fig1}-\ref{fig5}  also the functions $\alpha_{t}$, $\gamma_{t}$,  $t\in[0,T]$,  $T=1$,  are chosen as constants.

For  model  (\ref{reserves}), (\ref{al1}), (\ref{al2}),  (\ref{alCIgeneral}),   (\ref{rho}) the loss distribution  of the banks   defaulted in  $[0,T]$,  $T=1$   (shown   with a dashed line in Figures   \ref{fig1}-\ref{fig5})  
 is a unimodal distribution with a unique maximum    corresponding to a maximizer 
(or to several adjacent maximizers)    located in the interior of the interval $[0,N]$, $N=10$. 
Instead   when  we consider  model (\ref{reserves}), (\ref{alcoupled1}), (\ref{alcoupled2}), (\ref{alcoupled3}), (\ref{logalCI}),   (\ref{rho})  the loss distribution of the banks   defaulted in  the time interval $[0,T]$,  $T=1$  (shown   with a solid line  in Figures     \ref{fig1}-\ref{fig5})  has a  bump   near zero defaults and   a bump  near $N$ defaults and is small  in between, that is  is a bimodal distribution  with  two   maxima corresponding to two maximizers
(or to two disjoint sets of adjacent  maximizers)  located  at the endpoints of the interval $[0,N]$, $N=10$.
This is due to the  action  of the  cooperation mechanism   among banks in model 
 (\ref{reserves}), (\ref{alcoupled1}), (\ref{alcoupled2}), (\ref{alcoupled3}), (\ref{logalCI}),   (\ref{rho}).
Moreover the comparison between Figure  \ref{fig1} and Figures \ref{fig4}, \ref{fig5}  shows that the presence of a non zero correlation 
(i.e.  $\rho_{a} \ne 0$, $\rho_{l}  = 0$)
between the stochastic differentials on the right hand side   of the  assets equations  of the banks  (Figures \ref{fig4}, \ref{fig5})   increases  substantially  
the probability of ``extreme''  systemic risk with respect  to the probability of  the same event   in the zero correlation case (i.e.  $\rho_{a} = \rho_{l}  = 0$) (Figure  \ref{fig1}). 
Similar   phenomena  appear  when volatility and   correlation   coefficient jumps are present in  the  liabilities  equations.

Figures \ref{fig1}-\ref{fig5}   show that in  a homogeneous  bank population  the cooperation  among banks  introduced in  model 
(\ref{reserves}),  (\ref{alcoupled1}), (\ref{alcoupled2}), (\ref{alcoupled3}), (\ref{logalCI}),   (\ref{rho})  reduces  
 the default  probability  of the individual  bank  when compared to the default  probability 
of     the individual bank in  model (\ref{reserves}), (\ref{al1}), (\ref{al2}),  (\ref{alCIgeneral}),   (\ref{rho}) 
 at the expenses of  
the  default probability   of  the entire  (or of almost the entire) banking  system  that  is  greater in model 
(\ref{reserves}),  (\ref{alcoupled1}), (\ref{alcoupled2}), (\ref{alcoupled3}), (\ref{logalCI}),   (\ref{rho})  than  in model (\ref{reserves}), (\ref{al1}), (\ref{al2}),  (\ref{alCIgeneral}),   (\ref{rho}). 
Moreover the comparison of Figure  \ref{fig1} and  Figures  \ref{fig4},  \ref{fig5}
shows that this effect is enhanced by the presence of   ``collective''  behaviours in the bank population
 (i.e. is enhanced when $\rho_{a}^{2}$, $\rho_{l}^{2}$ are greater than zero). 
This is in agreement  with the findings of  \cite{Fouque2}, \cite{nostro1}, \cite{Haldane}, \cite{May},  
where it is shown that   for the stability of a    banking  system  an  excessive homogeneity  of the  bank population     is undesirable.
%

\section{The mean field approximation and the  pseudo mean field  approximation}\label{sec4}

For a survey of  the  mean field  approximation  in the context of statistical mechanics,   see, for example,  \cite{Gallavotti}, and the references therein. We limit our attention to the use of some ideas  taken from  the mean field  approximation   of statistical mechanics in the study  of the  banking system  models considered in the previous Sections.

Let us  begin considering      the mean field approximation of the  banking system  model (\ref{reserves}),  (\ref{alcoupled1}), (\ref{alcoupled2}), (\ref{alcoupled3}), (\ref{logalCI}),   (\ref{rho}).
When    the  stochastic differentials  of the 
equations (\ref{alcoupled2}), (\ref{alcoupled3})   $dW_{t}^{i}$,   $dZ_{t}^{i}$, $  t \ge 0$,   $i=1,2,\ldots,N$,  are independent, that is  when in   (\ref{rho}) we have:   $\rho_{a}^{2}=\rho_{a,t}^{2}=0$,  $\rho_{l}^{2}=\rho_{l,t}^{2}=0$, $t \ge 0$, 
 so that  in  (\ref{rhoa}),  (\ref{rhol})  we have:  $dW_{t}^{i}= d\widetilde{W}_{t}^{i}$,  $dZ_{t}^{i}= d\widetilde{Z}_{t}^{i}$, $  t \ge 0$,   $i=1,2,\ldots,N$,    the mean field approximation of the banking system  model (\ref{reserves}),  (\ref{alcoupled1}), (\ref{alcoupled2}), (\ref{alcoupled3}), (\ref{logalCI}),  (\ref{rho})  can be  deduced   proceeding as done in 
\cite{Fouque1},  \cite{nostro1}.
In fact   when  $\rho_{a}^{2}=\rho_{a,t}^{2}=0$,  $\rho_{l}^{2}=\rho_{l,t}^{2}=0$, $t \ge 0$,    and $N$   goes to infinity,     it is easy to see that  the mean  field limit   of    (\ref{reserves}),  (\ref{alcoupled1}), (\ref{alcoupled2}), (\ref{alcoupled3}), (\ref{logalCI}), (\ref{rho})  is given by:
\begin{eqnarray}
{\cal Y}_{t}={\cal A}_{t}-{\cal L}_{t} , \quad t \ge 0,  \label{MFg}
\end{eqnarray}
where
\begin{eqnarray}
{\cal A}_{t}=e^{{\cal G}_{t}}, \quad {\cal L}_{t}=e^{{\cal H}_{t}},  \quad  t \ge 0, \label{MFal1}
\end{eqnarray}
and ${\cal G}_{t}$, ${\cal H}_{t}$, $t \ge 0$, satisfy  the stochastic differential equations:
\begin{eqnarray}
&&   d \left({\cal G}_{t}   -  \ln(\varphi_{t})\right)= \alpha_{t} \left(\ln(\varphi_{t})   -{\cal G}_{t}   \right) dt+ \sigma_{a} dP_{t},\quad t>0, \label{MFal2}\\
&&   d \left({\cal H}_{t}   -  \ln(\phi_{t}) \right)= \gamma_{t} \left(\ln(\phi_{t})   -{\cal H}_{t}   \right) dt+ \sigma_{l} dQ_{t},\quad  t>0,   \label{MFal3}
\end{eqnarray}
with the  initial conditions:
\begin{equation}\label{MFalCI}
{\cal G}_{0} =\ln(\varphi_{0}), \quad {\cal H}_{0} =\ln(\phi_{0}).
\end{equation}
The stochastic processes  $P_{t}$, $Q_{t}$,  $ t \ge 0$, of (\ref{MFal2}),  (\ref{MFal3})  are standard Wiener processes such that $P_{0}=0$, $Q_{0}=0$,  $dP_{t},$  $dQ_{t},$   $t>0$, are their 
stochastic differentials and we have: 
\begin{equation}\label{MFrho}
\E(dP_{t} dQ_{t})=0, \quad   t>0.
\end{equation}

In the  mean field approximation  (\ref{MFg}),  (\ref{MFal1}), (\ref{MFal2}), (\ref{MFal3}), (\ref{MFalCI}),   (\ref{MFrho})  the   stochastic process   ${\cal Y}_{t}$, $t>0$,   represents  the capital  reserves of the   ``mean bank''  at time $t \ge 0$.  Similarly    the  stochastic processes ${\cal A}_{t}$,   ${\cal L}_{t}$, $t>0$,    represent,  respectively,   the assets  and  the liabilities of  the   ``mean bank''  at time  $t \ge 0$. 
Due to the homogeneity of the bank population,
when   $N$  goes to infinity  the assets, the liabilities and the  capital reserves of   the   banks  of  model  (\ref{reserves}),  (\ref{alcoupled1}), (\ref{alcoupled2}), (\ref{alcoupled3}), (\ref{logalCI}),  (\ref{rho})      behave, respectively,  like the assets, the liabilities and  the capital reserves of  the  ``mean bank'', that is behave like the  stochastic processes defined in     (\ref{MFal1}),  (\ref{MFg}).

%
%


Let us consider  the  banking system
  model (\ref{reserves}),  (\ref{alcoupled1}), (\ref{alcoupled2}), (\ref{alcoupled3}), (\ref{logalCI}),   (\ref{rho}) when the stochastic differentials
of  equations (\ref{alcoupled2}), (\ref{alcoupled3}) are correlated, that is when   $\rho_{a}$,  $\rho_{l}$ are non zero constants. 
Also in this  case  it is not difficult  to deduce the mean field approximation of the  banking system  model  (see, for example, \cite{Fouque2}), however, for later convenience,  
we prefer to introduce a  heuristic approximation  of   model (\ref{reserves}),  (\ref{alcoupled1}), (\ref{alcoupled2}), (\ref{alcoupled3}), (\ref{logalCI}),   (\ref{rho})    in the limit  $N$   goes to infinity  
that we call  pseudo mean field approximation that will be used  in Sections \ref{sec5} and \ref{sec6}     to govern   the probability of systemic risk     in a bounded time interval.
 In the  pseudo mean field  approximation     of the  banking system  model (\ref{reserves}),  (\ref{alcoupled1}), (\ref{alcoupled2}), (\ref{alcoupled3}), (\ref{logalCI}),  
 (\ref{rho})  the equations  (\ref{MFal2}), (\ref{MFal3}) are substituted, respectively,  with  the  equations:
\begin{eqnarray}
&&  d \left({\cal G}_{t}   -  \ln(\varphi_{t})\right)= 
 \alpha_{t} \left(  1- |\rho_{a}|   \right) \left(\ln(\varphi_{t})   -{\cal G}_{t}\right)  dt +\nonumber \\
 &&\hskip3.2truecm
g_{t}   \,   |\rho_{a} |    \left( \ln(\phi_{t}) - {\cal H}_{t} \right) dt+ \sigma_{a} dP_{t}, \quad   t>0, \label{MFal2corr}\\
&&   d \left({\cal H}_{t}   -  \ln(\phi_{t}) \right)=
\gamma_{t}  \left(  1- |\rho_{l}|  \right) \left(\ln(\phi_{t})   -{\cal H}_{t}   \right) dt+  \nonumber \\
&&\hskip3.2truecm
h_{t}  \,   |\rho_{l}| \left( \ln(\varphi_{t}) - {\cal G}_{t} \right) dt+ \sigma_{l} dQ_{t},\quad  t>0.  \label{MFal3corr}
\end{eqnarray}
The equations  (\ref{MFal2corr}), (\ref{MFal3corr})  are equipped with the  initial conditions  (\ref{MFalCI})  and the assumption  (\ref{MFrho}).
The  functions  $ g_{t} \ge 0 $, $ h_{t} \ge 0$, $t\ge0$,  are non negative functions  that will be  chosen later.   The  pseudo mean field  approximation   is completed adding the equations   (\ref{MFg}),  (\ref{MFal1}) to the equations  (\ref{MFal2corr}), (\ref{MFal3corr}), (\ref{MFalCI}),   (\ref{MFrho}).
In  the   pseudo mean field  approximation  (\ref{MFg}),  (\ref{MFal1}), (\ref{MFal2corr}), (\ref{MFal3corr}), (\ref{MFalCI}),   (\ref{MFrho})  the  stochastic processes ${\cal Y}_{t}$,   ${\cal A}_{t}$,   ${\cal L}_{t}$,  $t \ge 0,$ have the same meaning than in the mean field approximation, that is they 
represent,  respectively,  the capital  reserves,  the assets  and  the liabilities of  the   ``pseudo  mean bank''   as functions of time. 
The  equations    (\ref{MFg}),  (\ref{MFal1}), (\ref{MFal2corr}), (\ref{MFal3corr}), (\ref{MFalCI}),   (\ref{MFrho})    define  the dynamics  of the ``pseudo  mean bank''. 

When  $N$   goes to infinity  and the functions  $ g_{t}$, $ h_{t}$, $t\ge0$, are chosen appropriately,   the   ``pseudo  mean bank''    behaviour    ``approximates''  the behaviour   of the  ``mean bank''  and as a consequence   
``approximates''  the  behaviour  of the banks of model  (\ref{reserves}),  (\ref{alcoupled1}), (\ref{alcoupled2}), (\ref{alcoupled3}), (\ref{logalCI}),   (\ref{rho}). 
The choice of (\ref{MFg}),  (\ref{MFal1}), (\ref{MFal2corr}), (\ref{MFal3corr}), (\ref{MFalCI}),   (\ref{MFrho})   and in particular the choice of (\ref{MFal2corr}), (\ref{MFal3corr})  as pseudo mean field  approximation   is motivated by the following facts. 
First of all when    the stochastic differentials  $dW_{t}^{i}$,   $dZ_{t}^{i}$, $  t>0$,   $i=1,2,\ldots,N$,  of  equations (\ref{alcoupled2}), (\ref{alcoupled3})   are independent, that is  when in (\ref{rho})  we have  $\rho_{a}^{2}=0$,  $\rho_{l}^{2}=0$, $  t>0$,   the 
pseudo mean field  approximation  (\ref{MFg}),  (\ref{MFal1}), (\ref{MFal2corr}), (\ref{MFal3corr}), (\ref{MFalCI}),   (\ref{MFrho})  coincides with   the mean field  approximation  (\ref{MFg}),  (\ref{MFal1}), (\ref{MFal2}), (\ref{MFal3}), (\ref{MFalCI}),   (\ref{MFrho}).
Moreover  when  in the  equations (\ref{alcoupled2}), (\ref{alcoupled3})  the   stochastic differentials   $d W_{t}^{i}$,   $  t>0$,   $i=1,2,\ldots,N$, and   $d Z_{t}^{i}$,   $  t>0$,   $i=1,2,\ldots,N$,    are 
totally  correlated, that is   when  we have   $|\rho_{a}|=1$,   $|\rho_{l}|=1$, and 
   we choose  $ g_{t}= 0$, $ h_{t}= 0$, $t>0$,  the  pseudo mean field   approximation
     (\ref{MFg}),  (\ref{MFal1}), (\ref{MFal2corr}), (\ref{MFal3corr}), (\ref{MFalCI}),   (\ref{MFrho}) ``coincides''  with the banking system model
(\ref{reserves}),  (\ref{alcoupled1}), (\ref{alcoupled2}), (\ref{alcoupled3}), (\ref{logalCI}),   (\ref{rho}), with     $|\rho_{a}|=1$,  $|\rho_{l}|=1$, $  t>0$. 
That is   the  pseudo mean field  approximation   is ``exact''.
%
In fact when  $ |\rho_{a}|  = 1$, $ |\rho_{l}|  = 1$, $  t>0$,   
the initial conditions  $G_{0}^{i}=\ln(\varphi_{0}),$ $H_{0}^{i}=\ln(\phi_{0}), $ $i=1,2,\ldots,N$, imply  that 
 the  cooperation mechanism among banks  present  in  (\ref{alcoupled2}) and  in (\ref{alcoupled3}) has no influence on  the banking system dynamics. 
In fact   the previous  choices  imply  that in (\ref{alcoupled2})
$\alpha_{t}$,   $  t\ge0$,  multiplies  the null term, that is   imply  that $\displaystyle \sum_{j=1}^{N} \left( G_{t}^{j} -G_{t}^{i}\right) dt=0$, $t>0$. Similarly    the previous  choices imply that   in (\ref{alcoupled3})   $\gamma_{t}$,   $  t\ge0$,  multiplies  the null term, that is  imply that   $\displaystyle \sum_{j=1}^{N} \left( H_{t}^{j} -H_{t}^{i}\right) dt=0$, $t>0$.  
In fact  when  $ |\rho_{a}|  = 1$, $ |\rho_{l}|  = 1$, $  t>0$,  and  $G_{0}^{i}=\ln(\varphi_{0}),$ $H_{0}^{i}=\ln(\phi_{0}), $ $i=1,2,\ldots,N$,
 all the banks of the model satisfy the same equation  and can be considered as  a ``unique'' bank repeated $N$ times. 
 Note that      the condition   $ |\rho_{a}|  = 1$, $ |\rho_{l}|  = 1$, $  t>0$,    implies that  the Wiener  processes  present in the equations  relative to the different banks of the model coincide,  that is 
$ dW_{t}^{i}$, $  dZ_{t}^{i}$, $ t>0$,     in  (\ref{alcoupled2}),  (\ref{alcoupled3})    do not depend on $i$,   $i=1,2,\ldots,N$.
In this case all the  banks of the banking system model  are replicated  exactly  by the pseudo mean field  approximation
    (\ref{MFg}),  (\ref{MFal1}), (\ref{MFal2corr}), (\ref{MFal3corr}), (\ref{MFalCI}),   (\ref{MFrho}) when $ |\rho_{a}|  = 1$, $ |\rho_{l}|  = 1$, $  t>0$,   and  we choose  $ g_{t}= 0$,  $ h_{t}= 0$, $t>0$.
%
When  in  the equations (\ref{alcoupled2}), (\ref{alcoupled3})  the stochastic differentials  $d W_{t}^{i}$,   $d Z_{t}^{i}$, $  t>0$,   $i=1,2,\ldots,N$,  are partially  correlated, that is  when in (\ref{rho}) we have $0<|\rho_{a}|<1$,  $0<|\rho_{l}|<1$,   $  t>0$,  choosing appropriately the functions
$ g_{t}$, $ h_{t}$, $t>0$, 
the  pseudo mean field  approximation  (\ref{MFg}),  (\ref{MFal1}), (\ref{MFal2corr}), (\ref{MFal3corr}), (\ref{MFalCI}),   (\ref{MFrho})   ``interpolates''   between the extreme cases  $\rho_{a}=0$, $\rho_{l}=0$, $  t>0$,   and $|\rho_{a}|=1$, $|\rho_{l}|=1$, $  t>0$.
 Finally in Section \ref{sec6} in the systemic risk governance  the  form chosen for the  equations (\ref{MFal2corr}), (\ref{MFal3corr}) will make possible the use of the polynomial identity principle to determine the functions 
$\alpha_{t}$, $\gamma_{t}$, $t\ge0$, that regulate the cooperation mechanism among banks.

In Section \ref{sec6} we explain   the choice of the functions   $\varphi_{t}$,  $\phi_{t}$, $\alpha_{t}$, $\gamma_{t}$, $ g_{t}$, $ h_{t}$, $t\ge0$,  that  is used    to govern the systemic risk probability in a bounded time interval of model    (\ref{reserves}),  (\ref{alcoupled1}), (\ref{alcoupled2}), (\ref{alcoupled3}), (\ref{logalCI}),   (\ref{rho}).

Note that when  $\alpha_{t} =0$, $\gamma_{t} =0$,  $t\ge0$, and the functions  $\varphi_{t}$, $\phi_{t}$, $t\ge0$,  are positive constants,  there is no cooperation  among banks and no  intervention of the  monetary authority in the banking system dynamics.  
In this case    in the  pseudo mean field   approximation  we choose $ g_{t}= 0$, $ h_{t}= 0$, $t\ge0$.


%



%
%

%
\section{An optimal control problem  for the pseudo mean field approximation}\label{sec5}
%

  Let us consider  an  optimal control problem for the  pseudo mean field  approximation  (\ref{MFg}),  (\ref{MFal1}), (\ref{MFal2corr}), (\ref{MFal3corr}), (\ref{MFalCI}),   (\ref{MFrho})  of the  banking system  model    (\ref{reserves}),  (\ref{alcoupled1}), (\ref{alcoupled2}), (\ref{alcoupled3}), (\ref{logalCI}),   (\ref{rho}) when $0 \le |\rho_{a}|<1$, $0 \le |\rho_{l}|<1$, $  t>0$. 
%
%
Let $n$ be a positive integer, $ \R$ be  the set of  real numbers, $ \R^{n}$ be  the $n$-dimensional real Euclidean space  and   $ \R^{+} $ be  the set of the positive real numbers.

Given the positive  functions     $\varphi_{t}$,  $\phi_{t}$, $t\ge0$,    we define:
\begin{eqnarray}
{\cal Z}_{t}   = {\cal G}_{t}   -  \ln(\varphi_{t}), \quad t \ge 0, \label{zeta}
 \end{eqnarray}
 and
 \begin{eqnarray}
{\cal S}_{t}   = {\cal H}_{t}   -  \ln(\phi_{t}), \quad t \ge 0, \label{esse}
 \end{eqnarray}
  equations (\ref{MFal2corr}), (\ref{MFal3corr}), (\ref{MFalCI})   can be rewritten as follows:
\begin{eqnarray}
&&  d{\cal Z}_{t}= \beta_{a} (t,{\cal Z}_{t}, {\cal S}_{t}) dt+ \sigma_{a} dP_{t},\,  \,  t>0, \ \label{GeneralMF1}\\
&& d{\cal S}_{t}= \beta_{l} (t, {\cal Z}_{t}, {\cal S}_{t}) dt+ \sigma_{l} dQ_{t},\,  \,  t>0, \ \label{GeneralMF2}\\
&& {\cal Z}_{0}= 0 , \quad   {\cal S}_{0}= 0 , \label{GeneralMFCI}
\end{eqnarray}
where  $ \beta_{a}:  \R^{+} \times \R^{2} \rightarrow \R$  and  $ \beta_{l}:  \R^{+} \times \R^{2} \rightarrow \R$  are given by:
\begin{eqnarray}
&&\hskip-1truecm\beta_{a}=\beta_{a,t}=\beta_{a}(t,{\cal Z},{\cal S})= -  \alpha_{t}   \left(  1- |\rho_{a}|   \right)  {\cal Z}
-  |\rho_{a}|  \, g_{t}   \, {\cal S}, \quad    ({\cal Z},{\cal S}) \in  \R^{2}, \,\,  t>0, \label{betaa}\\
&&\hskip-1truecm \beta_{l}=\beta_{l,t}=\beta_{l}(t,{\cal Z},{\cal S})= - \gamma_{t}   \left(  1- |\rho_{l}|   \right)  {\cal S}
-  |\rho_{l} | \,  h_{t}   \, {\cal Z}, \quad ({\cal Z},{\cal S}) \in  \R^{2}, \,\,   t>0, \label{betal}
\end{eqnarray}
and in (\ref{GeneralMFCI}) the symbol  $0$ denotes the random variable concentrated in zero with probability one.

To choose  the functions   $\alpha_{t}$, $\gamma_{t}$, $ g_{t}$, $ h_{t}$, $t>0$, of  (\ref{betaa}),    (\ref{betal})   as done in  the systemic risk governance of Section \ref{sec6}  we  begin solving  the    stochastic optimal control   problem that follows.

Let  $T_{1}>0$ be a real number,  $\lambdavect=(\lambda_{1},\lambda_{2},\lambda_{3},\lambda_{4})\in \R^{4}$,   $\lambda_{i}>0$, $i=1,2,3,4$,   and $\br$ be the set of the real square integrable  stochastic processes defined in  $[0,T_{1}]$, that is a real stochastic process $\zeta=\zeta_{t}, $ $t\in [0,T_{1}]$, belongs to $\br$ if and only if $\displaystyle  \E \left(  \int_{0}^{T_{1}}   \zeta_{t}^{2} \, dt \right) <  +\infty$.
 We consider  the  following   stochastic optimal control   problem:
\begin{eqnarray}
&&  \min_{\beta_{1}, \beta_{2} \in \br}  U_{\lambdavect} (\beta_{1}, \beta_{2}) , 
  \label{Func1}
  \end{eqnarray}
  where  
  \begin{eqnarray}
&& \hskip-0.7truecm U_{\lambdavect} (\beta_{1}, \beta_{2}) =   \nonumber \\[3mm]
&& \hskip-0.7truecm
 \E     \left( \int_{0}^{T_{1}} \left[ |\rho_{a} \rho_{l} |({\cal Z}_{t}-{\cal S}_{t})^{2} +\lambda_{1} \beta_{1,t}^{2} +\lambda_{2} \beta_{2,t}^{2} +\lambda_{3} (1-|\rho_{a}|) {\cal Z}_{t}^{2}  +\lambda_{4} (1-|\rho_{l}|) {\cal S}_{t}^{2}     \right] dt  \right) , \nonumber \\
&&  \hskip2truecm  \beta_{1}, \,  \beta_{2} \in \br , \,  0 \le |\rho_{a}|<1, \, 0 \le |\rho_{l}|<1, \, \
  \label{Func2}
  \end{eqnarray}
  subject to
   \begin{eqnarray} 
&&        d{\cal Z}_{t}   =\beta_{1} \, dt +\sigma_{a} dP_{t},    \quad   t\in [0,T_{1}],   \label{Func3}     \\
&&        d{\cal S}_{t}   =\beta_{2} \, dt +\sigma_{l} dQ_{t},    \quad   t\in [0,T_{1}],   \label{Func4}     \\
&&        {\cal Z}_{0}   =0,    \quad  {\cal S}_{0}   =0.   \label{Func5}    
\end{eqnarray}
In the control problem   (\ref{Func1}), (\ref{Func2}), (\ref{Func3}), (\ref{Func4}),  (\ref{Func5})  the function   
$U_{\lambdavect} (\beta_{1}, \beta_{2})$ is the utility function,  
   $ \beta_{1}=\beta_{1,t}=\beta_{1}(t,{\cal Z}_{t},{\cal S}_{t}),  $  $ \beta_{2}=\beta_{2,t}=\beta_{2}(t,{\cal Z}_{t},{\cal S}_{t}), $ $t\in [0,T_{1}], $  are the control variables and  ${\cal Z}_{t}$, ${\cal S}_{t}$, $t\in [0,T_{1}], $ are the state variables. The random variables on the right hand side of 
   equation  (\ref{Func5})  must be interpreted as already done for  those of equation (\ref{GeneralMFCI}).

When $0<|\rho_{a}|<1$, $0<|\rho_{l}|<1$,    minimizing  the  utility function 
$U_{\lambdavect} (\beta_{1}, \beta_{2})$, 
$ \beta_{1}=\beta_{1,t}, $  $ \beta_{2}=\beta_{2,t}, $ $t\in [0,T_{1}],  $   
defined in   (\ref{Func2}) means  making small in the time interval $[0,T_{1}]$ the following quantities: 
${\it i)}$ the difference between  the capital reserves  of the  ``pseudo  mean bank''  and the capital reserves of the ``ideal bank'' ${\cal Z}_{t}-{\cal S}_{t}=({\cal G}_{t}   -  \ln(\varphi_{t}))-({\cal H}_{t}   -  \ln(\phi_{t}))$,   $t\in [0,T_{1}]$,  
${\it ii)}$     the  ``size'' of the  control variable $\beta_{1,t}$, $t\in [0,T_{1}]$,  
${\it iii)}$    the  ``size'' of the  control variable $\beta_{2,t}$, $t\in [0,T_{1}]$, 
${\it iv)}$    the ``size'' of ${\cal Z}_{t}$,  $t\in [0,T_{1}]$  (and therefore  the difference between   ${\cal G}_{t}$,  $t\in [0,T_{1}]$,   and the function  $ \ln(\varphi_{t})$, $t\in [0,T_{1}]$), 
  ${\it v)}$   the  ``size''  of ${\cal S}_{t}$,  $t\in [0,T_{1}]$  (and therefore  the difference between   ${\cal H}_{t}$,  $t\in [0,T_{1}]$,   and the function  $ \ln(\phi_{t})$, $t\in [0,T_{1}]$).
 These five goals correspond, respectively, to  making small the  addenda:  
 {\it i})  $\displaystyle \E      \left( \int_{0}^{T_{1}}  |\rho_{a} \rho_{l} |  ({\cal Z}_{t}-{\cal S}_{t})^{2} dt \right)$,   
  {\it ii}) $\displaystyle \E      \left( \int_{0}^{T_{1}} \lambda_{1} \beta_{1,t} ^{2} \,  dt  \right)$,
  {\it iii}) $\displaystyle \E      \left( \int_{0}^{T_{1}} \lambda_{2} \beta_{2,t} ^{2} \,  dt  \right) $,
  {\it iv}) $\displaystyle \E      \left( \int_{0}^{T_{1}} \lambda_{3} (1-|\rho_{a}|)  {\cal Z}_{t}^{2}  \,  dt  \right)$,
 {\it v}) $\displaystyle \E      \left( \int_{0}^{T_{1}} \lambda_{4}  (1-|\rho_{l}|)\right.$ $ \left.{\cal S}_{t}^{2} \,  dt  \right) $
of   the utility function $U_{\lambdavect}$ defined in   (\ref{Func2}).

 Note that when $\rho_{a}= 0$ and/or $ \rho_{l}=0$  the term 
 $\displaystyle \E      \left( \int_{0}^{T_{1}}  |\rho_{a} \rho_{l} |  ({\cal Z}_{t}-{\cal S}_{t})^{2} dt \right)$ 
 of $U_{\lambdavect} $   is   zero and  that in this case  minimizing   the  utility function 
$U_{\lambdavect} $
  corresponds    to pursuing only  four  of the five goals  listed above, that is  corresponds  to   pursuing
 the   goals of making small in the time interval $[0,T_{1}]$  the  quantities   ${\it ii)}$, ${\it iii)}$, ${\it iv)}$, ${\it v)}$. 

The control problem   (\ref{Func1}), (\ref{Func2}), (\ref{Func3}), (\ref{Func4}),  (\ref{Func5})  is a  linear-quadratic optimal control problem  (see \cite{Kolosov}).  Following  Kalman  \cite{Kolosov}  we assume  that  its value function    is a quadratic form in the  real variables  ${\cal Z}$, ${\cal S}$   with time dependent coefficients.  We have:

\noindent
{ \bf Proposition 1.}  Under the previous assumptions when $ 0 \le |\rho_{a}|<1$ and $0 \le |\rho_{l}|<1$ the optimal control  $\beta_{1}=\beta_{1,t}$, $\beta_{2}=\beta_{2,t}$, $t\in [0,T_{1}]$,  solution of  problem (\ref{Func1}), (\ref{Func2}), (\ref{Func3}), (\ref{Func4}), (\ref{Func5})    is  given by:
\begin{eqnarray}
&&\beta_{1}=\beta_{1,t}=\beta_{1}(t, {\cal Z}_{t}, {\cal S}_{t})= - \frac{1}{2\lambda_{1}} \left( 2 a(t)  {\cal Z}_{t} + c(t)  {\cal S}_{t} \right), \quad   t\in [0,T_{1}] ,  \label{soluzPropBeta1} \\
&&\beta_{2}=\beta_{1,t}=\beta_{2}(t, {\cal Z}_{t}, {\cal S}_{t})= - \frac{1}{2\lambda_{2}} \left( 2 b(t)  {\cal S}_{t} + c(t)  {\cal Z}_{t} \right), \quad   t\in [0,T_{1}] , \label{soluzPropBeta2} 
 \end{eqnarray}
  where ${\cal Z}_{t}$,  ${\cal S}_{t}$,  $t\in [0,T_{1}]$, are solution of the initial value problem   (\ref{Func3}), (\ref{Func4}),  (\ref{Func5}). 
The functions $a(t)$, $b(t)$, $c(t)$, $d(t)$,  $  t\in [0,T_{1}],$ are defined by   the following   final value problem:
\begin{eqnarray}
&&    \frac{\partial  a} {\partial t} = \frac{a^{2} } {\lambda_{1}}+ \frac{c^{2} } {4 \lambda_{2}}  -  |\rho_{a} \rho_{l} |   -\lambda_{3} (1-|\rho_{a}|) , \quad   \quad  t\in [0,T_{1}] , \hskip1truecm a(T_{1})=0,  \label{Riccati1}  \\[2mm]
&&    \frac{\partial b} {\partial t} = \frac{b^{2} } {\lambda_{2}}+ \frac{c^{2} } {4 \lambda_{1}}  -  |\rho_{a} \rho_{l} |   -\lambda_{4} (1-|\rho_{l}|) , \quad   \quad  t\in [0,T_{1}] , \hskip1.1truecm b(T_{1})=0,  \label{Riccati2}  \\[2mm]
&&    \frac{\partial  c} {\partial t} = \frac{ a c  } {\lambda_{1}}+ \frac{ b c  } {\lambda_{2}} +2 |\rho_{a} \rho_{l} | , \hskip3.3truecm  t\in [0,T_{1}] , \hskip1.1truecm c(T_{1})=0,  \label{Riccati3}  \\[2mm]
&&    \frac{\partial d } {\partial t} =  -a \sigma_{a}^{2} -  b \sigma_{l}^{2}, \hskip4.3truecm   t\in [0,T_{1}] , \hskip1.1truecm d(T_{1})=0. \label{Riccati4}  
 \end{eqnarray}

Note that the optimal control  (\ref{soluzPropBeta1}), (\ref{soluzPropBeta2}) does not depend from   the function $d(t)$, $  t\in [0,T_{1}]$ that appears
 in (\ref{Riccati4}); 
it depends only from   the functions  $a(t)$, $b(t)$, $c(t)$,  $  t\in [0,T_{1}],$ and  that the final value problem 
 (\ref{Riccati1}), (\ref{Riccati2}), (\ref{Riccati3}) satisfied by $a(t)$, $b(t)$, $c(t)$,  $  t\in [0,T_{1}],$ can be solved  independently  from 
 the final value problem  (\ref{Riccati4}) satisfied by  $d(t)$,  $  t\in [0,T_{1}]$. 
However the function $d(t)$,  $  t\in [0,T_{1}]$ is necessary 
 to   define   the value function  $V$ (see   (\ref{soluzProp}))  of the control problem (\ref{Func1}), (\ref{Func2}), (\ref{Func3}), (\ref{Func4}),  (\ref{Func5})  
and   the formulae  (\ref{soluzPropBeta1}), (\ref{soluzPropBeta2}) for   the  optimal control are deduced from the expression of the value function.

\noindent
{\bf Proof.} 
Let us   use    the  dynamic programming principle (see  \cite{Kolosov}) to solve  the control problem (\ref{Func1}), (\ref{Func2}), (\ref{Func3}), (\ref{Func4}), (\ref{Func5}).  That is let
\begin{eqnarray}
&& \hskip-1truecm  V(t,{\cal Z},{\cal S})= \min_{\beta_{1}, \beta_{2} \in \br}  \E     \left( \int_{t}^{T_{1}} \left[ |\rho_{a} \rho_{l} |({\cal Z}_{\tau}-{\cal S}_{\tau})^{2} +\lambda_{1} \beta_{1,\tau}^{2} +\lambda_{2} \beta_{2,\tau}^{2} +\lambda_{3} (1-|\rho_{a}|) {\cal Z}_{\tau}^{2}  +\right. \right. \nonumber\\
&& \hskip2truecm \left.  \left. \lambda_{4} (1-|\rho_{l}|) {\cal S}_{\tau}^{2}     \right] d\tau   \Big| 
  {\cal Z}_{t}={\cal Z},   {\cal S}_{t}={\cal S}\right) , \, \,   ({\cal Z} , {\cal S}) \in \R^{2},  \, t\in [0,T_{1}],   \label{valuefunction}
\end{eqnarray}
 be the value function  of the  control problem   (\ref{Func1}), (\ref{Func2}), (\ref{Func3}), (\ref{Func4}),  (\ref{Func5}).
 The  function $V(t,{\cal Z},{\cal S})$, $({\cal Z} , {\cal S}) \in \R^{2}, $  $  t\in [0,T_{1}]$, 
 satisfies the following Hamilton, Jacobi, Bellman equation (see \cite{Kolosov}):
\begin{eqnarray}
&& \frac{\partial } {\partial t} V(t,{\cal Z},{\cal S}) +
 \frac12 \sigma_{a}^{2}   \frac{\partial^{2} } {\partial {\cal Z}^{2}} V(t,{\cal Z},{\cal S}) +
  \frac12 \sigma_{l}^{2}   \frac{\partial^{2} } {\partial {\cal S}^{2}} V(t,{\cal Z},{\cal S}) +
|\rho_{a} \rho_{l} | \left(  {\cal Z}-  {\cal S} \right)^{2} +\nonumber\\
&&\lambda_{3} (1-|\rho_{a}|) {\cal Z}^{2} +\lambda_{4} (1-|\rho_{l}|) {\cal S}^{2} + 
  \calH\left( \frac{\partial } {\partial {\cal Z}} V(t,{\cal Z},{\cal S}), \frac{\partial } {\partial {\cal S}} V(t,{\cal Z},{\cal S}) \right)=0 ,
\nonumber  \\
&&     \hskip7truecm  ({\cal Z} , {\cal S}) \in \R^{2},  \, t\in [0,T_{1}],    \label{HJB}
\end{eqnarray}
with  final condition:
\begin{eqnarray}
&& V(T_{1},{\cal Z},{\cal S})=0 , \quad  ({\cal Z} , {\cal S}) \in \R^{2} , 
  \label{HJBCI}  
\end{eqnarray}
where
\begin{eqnarray}  \label{Hamilt} 
 \calH(p_{1},p_{2})= \min_{(\delta_{1},\delta_{2}) \in \R^{2}} \left( \delta_{1} p_{1}    + \lambda_{1}  \delta_{1}^{2} + \delta_{2} p_{2}    + \lambda_{2}  \delta_{2}^{2}\right)=-\frac{p_{1}^{2}}{4 \lambda_{1}}-\frac{p_{2}^{2}}{4 \lambda_{2}},  \quad (p_{1}, p_{2}) \in \R^{2},  
\end{eqnarray}
is the Hamiltonian function  of the optimal control problem   (\ref{Func1}), (\ref{Func2}), (\ref{Func3}), (\ref{Func4}),  (\ref{Func5}).\\
Using    (\ref{Hamilt})  equation  (\ref{HJB}) becomes:
\begin{eqnarray}
&& \frac{\partial } {\partial t} V(t,{\cal Z},{\cal S}) +
 \frac12 \sigma_{a}^{2}   \frac{\partial^{2} } {\partial {\cal Z}^{2}} V(t,{\cal Z},{\cal S}) +
  \frac12 \sigma_{l}^{2}   \frac{\partial^{2} } {\partial {\cal S}^{2}} V(t,{\cal Z},{\cal S}) +
|\rho_{a} \rho_{l} | \left(  {\cal Z}-  {\cal S} \right)^{2} +\nonumber\\
&&\quad \quad\lambda_{3} (1-|\rho_{a}|) {\cal Z}^{2} +\lambda_{4} (1-|\rho_{l}|) {\cal S}^{2} 
 -\frac{1}{4 \lambda_{1}}\left( \frac{\partial } {\partial {\cal Z}} V(t,{\cal Z},{\cal S}) \right)^{2}-\nonumber\\
 &&\quad \quad\frac{1}{4 \lambda_{2}}\left( \frac{\partial } {\partial {\cal S}} V(t,{\cal Z},{\cal S}) \right)^{2}
=0 , \quad ({\cal Z} , {\cal S}) \in \R^{2},  \, t\in [0,T_{1}],    \label{HJB2}
\end{eqnarray}
with the final condition (\ref{HJBCI}).\\
Following    Kalman \cite{Kolosov}  we assume  that the value function    solution of problem  (\ref{HJB2}),  (\ref{HJBCI})  is of   the form:
\begin{eqnarray} \label{soluzProp} 
V(t,{\cal Z},{\cal S})=a(t)  {\cal Z}^{2}  +b(t) {\cal S}^{2}  +c(t){\cal Z} {\cal S} + d(t),     \quad  \quad ({\cal Z} , {\cal S}) \in \R^{2}, \, \,  t\in [0,T_{1}] ,
 \end{eqnarray}
where   $a(t)$,  $b(t)$, $c(t)$,  $d(t)$, $ t\in [0,T_{1}],$  are   functions  to be determined.\\
 Substituting  (\ref{soluzProp})  in  (\ref{HJB2}),  (\ref{HJBCI}) and  using the  polynomial identity principle   it is easy to see  that the final value   problem  for the  Hamilton, Jacobi, Bellman equation  (\ref{HJB2}),  (\ref{HJBCI})  reduces to   the final value problem  (\ref{Riccati1}), (\ref{Riccati2}), (\ref{Riccati3}), (\ref{Riccati4}).

Problem (\ref{Riccati1}), (\ref{Riccati2}), (\ref{Riccati3}), (\ref{Riccati4}) is a  final value problem for a  system of  Riccati  ordinary differential equations.  In general systems  of this kind   have only local solutions. This means that, in general,  a solution  of  (\ref{Riccati1}), (\ref{Riccati2}), (\ref{Riccati3}), (\ref{Riccati4})  in the time interval $[0,T_{1}]$   may not exist. When  this is the  case the assumption    (\ref{soluzProp})   about  the form of the    value function  is not good enough to solve problem  (\ref{Func1}), (\ref{Func2}), (\ref{Func3}), (\ref{Func4}), (\ref{Func5}) and  we do not go  any further  in the study of the control problem    (\ref{Func1}), (\ref{Func2}), (\ref{Func3}), (\ref{Func4}), (\ref{Func5}).
From now on we assume that the final value   problem (\ref{Riccati1}), (\ref{Riccati2}), (\ref{Riccati3}), (\ref{Riccati4}) has a solution defined in $[0,T_{1}]$.
 
 From  the knowledge of the  value function $V$  defined in  (\ref{soluzProp}) solution of   (\ref{HJB2}),  (\ref{HJBCI})   the optimal control  $\beta_{1}=\beta_{1,t}$,  $\beta_{2}=\beta_{2,t}$, $  t\in [0,T_{1}], $  solution of (\ref{Func1}), (\ref{Func2}), (\ref{Func3}), (\ref{Func4}),  (\ref{Func5}) is determined  using the  formulae:
\begin{eqnarray} 
&& \hskip-0.5truecm
\beta_{1}=\beta_{1,t}=\beta_{1}(t,{\cal Z}_{t},{\cal S}_{t})= - \frac{1}{2\lambda_{1}} \frac{\partial } {\partial {\cal Z}} V(t,{\cal Z},{\cal S})
\Big|_{{\cal Z}={\cal Z}_{t},{\cal S}={\cal S}_{t}}= - \frac{1}{2\lambda_{1}} \left( 2 a(t)  {\cal Z}_{t} + c(t)  {\cal S}_{t} \right), \nonumber\\[2mm]
&& \hskip8truecm  t\in [0,T_{1}] , \label{soluz1} \\[2mm]
&&  \hskip-0.5truecm
\beta_{2}=\beta_{2,t}=\beta_{2}(t,{\cal Z}_{t},{\cal S}_{t})= - \frac{1}{2\lambda_{2}} \frac{\partial } {\partial {\cal S}} V(t,{\cal Z},{\cal S})
\Big|_{{\cal Z}={\cal Z}_{t},{\cal S}={\cal S}_{t}}= - \frac{1}{2\lambda_{2}} \left( 2 b(t)  {\cal S}_{t} + c(t)  {\cal Z}_{t} \right), \nonumber\\[2mm]
&& \hskip8truecm   t\in [0,T_{1}] , \label{soluz2} 
 \end{eqnarray}
where ${\cal Z}_{t}$,  ${\cal S}_{t}$, $t\in [0,T_{1}]$,  are the solution of (\ref{Func3}), (\ref{Func4}),  (\ref{Func5}) when $\beta_{1}$=$\beta_{1,t}$,  $\beta_{2}$=$\beta_{2,t}$,  $  t\in [0,T_{1}], $   are given by  (\ref{soluz1}), (\ref{soluz2}).\\

Problem (\ref{Func1}), (\ref{Func2}), (\ref{Func3}), (\ref{Func4}), (\ref{Func5})   is the optimal control  problem   used  to govern the  pseudo mean field  approximation  (\ref{MFg}),  (\ref{MFal1}), (\ref{MFal2corr}), (\ref{MFal3corr}), (\ref{MFalCI}),   (\ref{MFrho})  of the  banking system  model   (\ref{reserves}),  (\ref{alcoupled1}), (\ref{alcoupled2}), (\ref{alcoupled3}), (\ref{logalCI}),   (\ref{rho}).
In fact,   when  $0 \le |\rho_{a}|<1$, $0 \le  |\rho_{l}|<1$, given the optimal  control $\beta_{1}$, $\beta_{2}$  defined  in (\ref{soluz1}), (\ref{soluz2})     we determine   the functions $\beta_{a}$, $\beta_{l}$ of  (\ref{betaa}),    (\ref{betal})   imposing  the identities $\beta_{a}({\cal Z} , {\cal S})=\beta_{1}({\cal Z} , {\cal S})$,  $\beta_{l}({\cal Z} , {\cal S})=\beta_{2}({\cal Z} , {\cal S})$, $({\cal Z} , {\cal S}) \in \R^{2},$  and  using the  polynomial identity principle  in the variables $({\cal Z} , {\cal S}) \in \R^{2}$. We have:
\begin{eqnarray} 
&&\displaystyle\alpha_{t}=\frac{a(t)} {\lambda_{1}(1- |\rho_{a}| ) },  \quad   \displaystyle\gamma_{t}=\frac{b(t)} {\lambda_{2}(1- |\rho_{l}| ) }, \quad 
  t\in [0,T_{1}], \quad
0 \le |\rho_{a}|,  |\rho_{l}|<1,  \label{alphagammaott} 
 \end{eqnarray}
and
\begin{eqnarray}  \label{alphagammaott2} 
\displaystyle g_{t}=\frac{c(t)} {2 \lambda_{1} \, |\rho_{a}|  },  \quad   \displaystyle  h_{t}=\frac{c(t)} {2 \lambda_{2} \, |\rho_{l}|  }  , \quad  t\in [0,T_{1}], \quad
0< |\rho_{a}|,  |\rho_{l}|<1,
 \end{eqnarray}
or   
\begin{eqnarray}  \label{alphagammaott3} 
&& h_{t}=0 \quad   \text{and/or}  \quad  g_{t}=0,    \quad  t\in [0,T_{1}], \quad
 \rho_{a}=  0   \quad   \text{and/or}  \quad  \rho_{l}=  0.
 \end{eqnarray}
Let us point out  that  when $\rho_{a}=  0$ and/or $ \rho_{l}=0$  the function $c(t)=0$, $t\in [0,T_{1}]$ is a solution of   (\ref{Riccati3}).
Moreover note that the use of the polynomial identity principle  in the deduction of  (\ref{alphagammaott}),  (\ref{alphagammaott2}),  (\ref{alphagammaott3}) 
is possible thanks to the form of equations (\ref{MFal2corr}), (\ref{MFal3corr}) of the pseudo mean field approximation.

Recall  that the function 
 $ \alpha_{t}$, $  t\in [0,T_{1}],$  defined in (\ref{alphagammaott}) 
  is a function   that  substituted in (\ref{alcoupled2})   induces  the trajectories of the  logarithm of the assets to swarm around $\ln(\varphi_{t})$, $  t\in [0,T_{1}],$ and therefore induces  the trajectories of the  assets to swarm around $\varphi_{t}$, $  t\in [0,T_{1}].$
 Similarly  the function 
 $ \gamma_{t}$, $  t\in [0,T_{1}],$   defined in (\ref{alphagammaott}) is a function that  substituted in (\ref{alcoupled3})   induces   the trajectories of the  logarithms of the liabilities to swarm around $\ln(\phi_{t})$, $  t\in [0,T_{1}],$ and therefore  induces  the trajectories of the  liabilities to swarm around $\phi_{t}$, $  t\in [0,T_{1}].$ 

Remember that in (\ref{alcoupled2}), (\ref{alcoupled3}) the constraints  $\alpha_{t} \ge 0, $ $\gamma_{t} \ge 0$,  $t\in [0,T_{1}]$,   must be satisfied. When   they are not  satisfied by the choices  of $\alpha_{t},$ $ \gamma_{t}$,  $t\in [0,T_{1}]$, made in  (\ref{alphagammaott}) they are enforced. 
In  the numerical experiments discussed  in Section  \ref{sec6}  when the functions $\alpha_{t}$ and/or  $ \gamma_{t}$, 
$t\in [0,T_{1}]$,  determined using  (\ref{alphagammaott}) are negative,  we choose   $\alpha_{t}=0$ and/or $\gamma_{t}=0$, $t\in [0,T_{1}]$.

Note that the formulae  (\ref{alphagammaott}),  (\ref{alphagammaott2}),  (\ref{alphagammaott3})  provide a choice  of the functions 
 $ \alpha_{t}$,  $ \gamma_{t}$, $ g_{t}$,  $ h_{t}$  when $  t\in [0,T_{1}]$;  to choose the functions  $ \alpha_{t}$,  $ \gamma_{t}$, $ g_{t}$,  $ h_{t}$
when  $t>0$ the previous formulae  must be adapted  to  take care of the repeated solution  of control problems  similar  to the one considered here.

%
\section{The systemic risk governance }\label{sec6}
%

Let  $T_{2}>0$ be a real number and  consider the problem of  governing    the probability    of   systemic risk in the  time interval $[0,T_{2}]$  in 
model   (\ref{reserves}),  (\ref{alcoupled1}), (\ref{alcoupled2}), (\ref{alcoupled3}), (\ref{logalCI}),   (\ref{rho})    in  absence or  in presence of   shocks acting on the banking system.
Given $ \tau_{1}, \tau_{2}$  such that $ 0\le \tau_{1}<\tau_{2} \le  T_{2}$,  and the  interval  $[\tau_{1},\tau_{2}] \subseteq [0,T_{2}]$,    
let us  consider the governance of    systemic risk    in the time interval $[\tau_{1},\tau_{2}]$.
The  goal of the governance    is to keep          the   probability of    systemic risk in the time interval $[\tau_{1},\tau_{2}]$,  $\pr (SR_{[\tau_{1},\tau_{2}]})$,  between two given thresholds. 
%
%
%
The   systemic risk governance  pursues its goal trying  to   keep the assets, the liabilities and  the capital reserves of the banks  of the model  ``close'', respectively, to  the  assets, the liabilities and  the 
capital reserves  of the ``ideal bank'',   that is close, respectively,  to    the   functions $ \varphi_{t}>0$, $ \phi_{t}>0$ and  $\xi_{t} =\varphi_{t}- \phi_{t}> 0$,  $t\in [\tau_{1},\tau_{2}]$.
Given    the choice   of the functions   $\varphi_{t}$,  $\phi_{t}$,  $\xi_{t}=\varphi_{t}- \phi_{t}$, $t\in  [\tau_{1},\tau_{2}]$,  the governance is based on the solution of   the   optimal control problem   (\ref{Func1}), (\ref{Func2}), (\ref{Func3}), (\ref{Func4}), (\ref{Func5})   and  on  its relation with 
 the banking system  model   (\ref{reserves}),  (\ref{alcoupled1}), (\ref{alcoupled2}), (\ref{alcoupled3}), (\ref{logalCI}),   (\ref{rho})  when the functions
  $\alpha_{t}$, $\gamma_{t}$, $  t\in  [\tau_{1},\tau_{2}], $  are chosen  adapting formula       (\ref{alphagammaott})
   deduced for   the    time interval $[0,T_{1}]$  to the time interval $ [\tau_{1},\tau_{2}]$.
In fact  the choice of  the functions $\alpha_{t}$, $\gamma_{t}$,  $t\in [\tau_{1},\tau_{2}]$,   obtained  adapting formula    (\ref{alphagammaott}) to the time interval $[\tau_{1},\tau_{2}], $  creates   a ``swarming''  effect  of the assets and of the liabilities of the banks of the model  around, respectively,  the functions $\varphi_{t}$,  $\phi_{t}$,  $t\in  [\tau_{1},\tau_{2}]$, and, as a consequence,  creates 
  a ``swarming''  effect    of  the capital reserves of the banks of the model  around the  function  $\xi_{t}$,  $t\in [\tau_{1},\tau_{2}]$.

We assume that  the  decisions about systemic  risk governance  in the time interval $[\tau_{1},\tau_{2}]$ are  taken at time $t=\tau_{1}$. 
 Going into details  to  pursue  the  goal  of  keeping  the   probability of    systemic risk in the time interval $[\tau_{1},\tau_{2}]$,  $\pr (SR_{[\tau_{1},\tau_{2}]})$,  between two given thresholds  the first thing to do at time $t=\tau_{1}$  is to choose appropriately        the functions   $\varphi_{t}$, $\phi_{t}$,    $\xi_{t}=\varphi_{t}- \phi_{t}$, $t\in [\tau_{1},\tau_{2}]$.
 In fact it is easy to see  that 
 increasing    $\xi_{t}>0$,   $t\in [\tau_{1},\tau_{2}]$,  the systemic risk  probability in $ [\tau_{1},\tau_{2}]$ decreases  and  that  decreasing   $\xi_{t}>0$, 
    $t\in [\tau_{1},\tau_{2}]$,  the systemic risk  probability in $[\tau_{1},\tau_{2}]$ increases.  
 Moreover,  since $\xi_{t} =\varphi_{t}- \phi_{t}>0$,  $t\in [\tau_{1},\tau_{2}]$,   increasing    $\xi_{t}$, $t\in [\tau_{1},\tau_{2}]$, can be done    increasing    $\varphi_{t}  $ leaving  unchanged  $\phi_{t}$,   $t\in [\tau_{1},\tau_{2}]$, or decreasing $\phi_{t}  $  leaving  unchanged  $\varphi_{t}$,   $t\in [\tau_{1},\tau_{2}]$,  or changing  at the same time   $\varphi_{t}  $ and  $\phi_{t}  $,   $t\in [\tau_{1},\tau_{2}]$.
Similarly   decreasing    $\xi_{t}  $, $t\in [\tau_{1},\tau_{2}]$,  can  be  done   either   decreasing    $\varphi_{t}  $ leaving  unchanged  $\phi_{t}$,   $t\in [\tau_{1},\tau_{2}]$, or increasing $\phi_{t}  $ leaving  unchanged  $\varphi_{t}$,   $t\in [\tau_{1},\tau_{2}]$, or changing  at the same time    $\varphi_{t}  $ and  $\phi_{t}  $,   $t\in [\tau_{1},\tau_{2}]$.
 
Given the thresholds $S_{1}$, $S_{2}$, such that  $0<S_{1}<S_{2}<1$, and  $\varphi_{\tau_{1}}$, $\phi_{\tau_{1}}$,   $\xi_{\tau_{1}}=\varphi_{\tau_{1}}-\phi_{\tau_{1}}$   we want
 to choose    the functions     $\varphi_{t}$, $\phi_{t}$,  $\xi_{t}=\varphi_{t}- \phi_{t}$, $t\in [\tau_{1},\tau_{2}]$,  such  that     the probability of    systemic risk  in the time interval $[\tau_{1},\tau_{2}]$  satisfies the following inequalities:
\begin{eqnarray} \label{striscia}
S_{1} \le \pr (SR_{[\tau_{1},\tau_{2}]}) \le S_{2}.
 \end{eqnarray}

We   define some simple rules  that  are used    to choose  the functions  
  $\varphi_{t}$, $\phi_{t}$,    $\xi_{t}=\varphi_{t}- \phi_{t}$, $t\in [\tau_{1},\tau_{2}]$
  in order to satisfy (\ref{striscia}).
  At time $t=\tau_{1}$   we start making  the  ``simplest''      possible choice   of    $\varphi_{t}$, $\phi_{t}$,    $\xi_{t}=\varphi_{t}- \phi_{t}$,  $t\in [\tau_{1},\tau_{2}]$,     that is  we choose:  $\varphi_{t}=\varphi_{\tau_{1}}$, 
$\phi_{t}=\phi_{\tau_{1}}$,  $\xi_{t}=\xi_{\tau_{1}}$,  $t\in [\tau_{1},\tau_{2}]$.  
 In correspondence to  this choice   the functions   $\alpha_{t}$, $\gamma_{t}$, $t\in [\tau_{1},\tau_{2}]$,   are determined  adapting  formula (\ref{alphagammaott})    to the time interval  $[\tau_{1},\tau_{2}]$  
and  the probability   of    systemic risk   in the time interval $[\tau_{1},\tau_{2}]$,  $\pr (SR_{[\tau_{1},\tau_{2}]})$, is evaluated using statistical simulation. 
Note that $\pr (SR_{[\tau_{1},\tau_{2}]})$ depends not only  from  the functions    $\varphi_{t}$, $\phi_{t}$, $\alpha_{t}$, $\gamma_{t}$, $t\in [\tau_{1},\tau_{2}]$, but also  from  the  random variables $a_{\tau_{1}}^{i}$, $l_{\tau_{1}}^{i}$, $c_{\tau_{1}}^{i}   =a_{\tau_{1}}^{i}-l_{\tau_{1}}^{i}$,  $  i=1,2,\ldots,N$.
Based on the value of $\pr (SR_{[\tau_{1},\tau_{2}]})$     the following actions are taken:
 \begin{enumerate} 
 \item[] {\it Strategy 1}:   if $\pr (SR_{[\tau_{1},\tau_{2}]}) > S_{2}$  the monetary authority    changes the  functions  
 $\varphi_{t}$, $\phi_{t}$,   $\xi_{t}=\varphi_{t}- \phi_{t}$, $t\in [\tau_{1}, \tau_{2}]$,   to ``swarm''  the trajectories   of the  capital reserves   of the  banking system model (\ref{reserves}),  (\ref{alcoupled1}), (\ref{alcoupled2}), (\ref{alcoupled3}), (\ref{logalCI}),   (\ref{rho})  ``upward'', 
  that is  the monetary authority   increases   $\xi_{t}>0$, $t\in [\tau_{1}, \tau_{2}]$. 
  This is  done  in   one of the following ways:
        \begin{enumerate} 
               \item[] {\it Strategy 1a}:    increasing    $\varphi_{t}>0$ leaving  unchanged  $\phi_{t}>0$,   $t\in [\tau_{1},\tau_{2}]$;
               \item[] {\it Strategy 1b}:    decreasing   $\phi_{t}>0$ leaving  unchanged  $\varphi_{t}>0$,   $t\in [\tau_{1},\tau_{2}]$;
               \item[] {\it Strategy 1c}:     changing both $\varphi_{t}>0$ and     $\phi_{t}>0$,    $t\in [\tau_{1},\tau_{2}]$.
        \end{enumerate}          
 \item[] {\it Strategy 2}:  if $\pr (SR_{[\tau_{1},\tau_{2}]}) < S_{1}$  the monetary authority    changes the   functions  
  $\varphi_{t}$, $\phi_{t}$,   $\xi_{t}=\varphi_{t}- \phi_{t}$,  $t\in [\tau_{1}, \tau_{2}]$,   to ``swarm''  the trajectories   of the  capital reserves of the  banking system model (\ref{reserves}),  (\ref{alcoupled1}), (\ref{alcoupled2}), (\ref{alcoupled3}), (\ref{logalCI}),   (\ref{rho})  ``downward'',  
  that is  the monetary authority  decreases  $\xi_{t}>0$, $t\in [\tau_{1}, \tau_{2}]$.  
     This is  done  in   one of the following ways:
        \begin{enumerate} 
                \item[] {\it Strategy 2a}:   decreasing    $\varphi_{t}>0$ leaving  unchanged  $\phi_{t}>0$,   $t\in [\tau_{1},\tau_{2}]$;
                \item[] {\it Strategy 2b}:    increasing   $\phi_{t}>0$  leaving  unchanged  $\varphi_{t}>0$,   $t\in [\tau_{1},\tau_{2}]$;
                 \item[] {\it Strategy 2c}:    changing both $\varphi_{t}>0$ and     $\phi_{t}>0$,    $t\in [\tau_{1},\tau_{2}]$.
        \end{enumerate}                 
 \item[] {\it Strategy 3}:  if  $S_{1} \le \pr (SR_{[\tau_{1},\tau_{2}]}) \le S_{2} $    the monetary authority    leaves the functions  
  $\varphi_{t}$, $\phi_{t}$,  $\xi_{t}=\varphi_{t}- \phi_{t}$, $t\in [\tau_{1}, \tau_{2}]$,    unchanged.
  \end{enumerate} 

  Note that  at  time  $t=\tau_{1}$  the monetary authority    makes  its decisions  about systemic risk governance in the  time interval $[\tau_{1}, \tau_{2}]$   assuming  that the volatilities   $\sigma_{a}$,  $\sigma_{l}$  and  the correlation coefficients $\rho_{a}^{2}$, $\rho_{l}^{2}$ in the  time interval $[\tau_{1}, \tau_{2}]$  remain  constant at  the  value that they have  at  time $ t= \tau_{1}$. That is the monetary authority  does not foresee     volatility and/or correlation shocks that hit the banking system in the  time interval $[\tau_{1}, \tau_{2}]$, simply reacts to them after   they have occurred.
 
The choice of acting on the assets $\varphi_{t}$,  $t\in [\tau_{1},\tau_{2}]$, or on the liabilities $\phi_{t}$,  $t\in [\tau_{1},\tau_{2}]$, of the  ``ideal bank'',    depends from the kind of   shock   that must be confronted. 
For example    in  presence of a   volatility   shock on the  side of the assets  occurred before $t=\tau_{1}$  (the  systemic risk governance decision time), 
 that is reacting to  a jump of the function $\sigma_{a}$ occurred before $t=\tau_{1}$,
  it  is  natural  at time  $ t= \tau_{1}$  to  increase/decrease  $\xi_{t}$,  $t\in [\tau_{1},\tau_{2}]$,    simply increasing/decreasing   $\varphi_{t}>0$,   $t\in [\tau_{1},\tau_{2}]$, 
leaving  unchanged  $\phi_{t}>0$,   $t\in [\tau_{1},\tau_{2}]$. In other words in this situation it is natural to limit 
 the actions  considered by  the monetary authority  to  {\it Strategy 1a},  {\it 2a} and  {\it  3}. 
 Similarly   in  presence of a      volatility   shock on the  side of the liabilities   occurred before $t=\tau_{1}$,
 that is reacting to the  presence of a jump of the function $\sigma_{l}$ occurred before $t=\tau_{1}$,
  it  is  natural    at time  $ t= \tau_{1}$   to  increase/to decrease  $\xi_{t}$,  $t\in [\tau_{1},\tau_{2}]$,   simply decreasing/increasing    $\phi_{t}>0$,   $t\in [\tau_{1},\tau_{2}]$, leaving  unchanged  $\varphi_{t}>0$,   $t\in [\tau_{1},\tau_{2}]$. That is  in this situation  it is natural to limit  the actions  considered by  the monetary authority    to   {\it Strategy 1b},  {\it 2b} and  {\it  3}.

When    possible the strategy of increasing the assets  of the ``ideal bank''    is  more desirable  for the  well being   of the  economy than the strategy of decreasing the liabilities  of the ``ideal bank''. 
In fact  increasing the assets  induces a similar behaviour   of the assets of the  banks of the banking system and this    keeps   the wheels of the economy turning, while   decreasing the liabilities   
induces a similar behaviour   of the   liabilities     of the  banks of the banking system and  has the  effect of slowing  down   the economy. Taking it to extremes,  when   possible, the   monetary authority should   prefer  {\it   Strategies 1a},  {\it 2b} and  {\it  3}  to  {\it   Strategies 1b},  {\it 1c}, {\it 2a}, {\it  2c}

The choice  between the  Strategies   {\it 1a}, {\it 1b}, {\it 1c},  or   {\it 2a}, {\it 2b}, {\it 2c}, is based on the comparison of  these  strategies
from the   systemic risk point of view.  
A possible criterion to compare   Strategies {\it 1a}, {\it 1b}, {\it 1c},  or   {\it 2a}, {\it 2b}, {\it 2c}  from the   systemic risk point of view is to evaluate 
the  corresponding  loss distributions  of the banks defaulted in the time interval  $ [\tau_{1},\tau_{2}]$.   The strategy    associated  to  the  loss distribution  with     the ``smallest tail'' must  be considered as the best strategy.
  For simplicity  we do not pursue this goal  here.

  
Let us discuss some numerical experiments  of   systemic risk governance. That is let us  present the results obtained   considering   the governance of   systemic risk   in the next year during a period of two years in   model   (\ref{reserves}),  (\ref{alcoupled1}), (\ref{alcoupled2}), (\ref{alcoupled3}), (\ref{logalCI}),   (\ref{rho}) in  absence or  in presence of  shocks acting on the banking system. Governance decisions are taken at the beginning of each quarter during the two years period studied.
 For the sake of simplicity  we consider only  the following  types of shocks:  volatility   shocks on the  side of the assets and volatility   shocks on the  side of the liabilities.  The occurrence of these shocks  is simulated  respectively  with  jumps of  the volatilities $\sigma_{a}$,  $\sigma_{l}$, of  the stochastic differential equations   of  the assets (\ref{alcoupled2})  and of the liabilities (\ref{alcoupled3}).  Note that together with  jumps in the  volatility coefficients    sometime we consider   jumps   in   the correlation coefficients   $\rho_{a}$,   $\rho_{l}$, of the  stochastic differentials   on the right  hand side of   equations  
  (\ref{alcoupled2}), (\ref{alcoupled3}).  
 Moreover  when   there are no shocks acting on the banking  system or  when the monetary authority  faces a   volatility   shock on the  side of the assets  of the banks we consider   as possible  only the  actions described in  {\it Strategy 1a},  {\it 2a},  {\it  3} and  in {\it Strategy 1a},  {\it  2b},  {\it  3}. 
Similarly  when  the monetary authority  faces a   volatility   shock on the  side of the liabilities  of the banks 
 we  consider   as possible  only the  actions described in {\it Strategy 1a},  {\it  2b},  {\it  3} and in  {\it Strategy 1b},  {\it  2b}, {\it  3}. 
 
 In the  experiments  we study  a banking system model  with $N=10$ banks with   a time horizon $T_{2}$ of three years, that is  we choose  the time unit equal to one year and $T_{2}=3$.
  We suppose that  governance  decisions are taken  quarterly, that is  the time step  of  the governance decisions   is  $\Delta \tau=1/4$.
In the time interval $[0,T_{2}]$ we consider   the time intervals $ [\tau_{1}^{j},\tau_{2}^{j}] \subset [0,T_{2}]$, $T_{2}=3$,   where $\displaystyle \tau_{1}^{j}=j \cdot \Delta \tau$ and $\tau_{2}^{j}=\tau_{1}^{j}+1$, $j=0,1,\ldots,8$, and  governance  decisions  are taken at the times $t=\displaystyle \tau_{1}^{j}$, $j=0,1,\ldots,8$. 
That is at time $t=\displaystyle \tau_{1}^{j}$ it  is taken the decision relative to  systemic risk in the time interval  $ [\tau_{1}^{j},\tau_{2}^{j}]$, 
  at time $t=\displaystyle \tau_{1}^{j}$  this   is the systemic risk in the next year,  $j=0,1,\ldots,8$.

 In  the time intervals $ [\tau_{1}^{j},\tau_{2}^{j}]$, $j=0,1,\ldots,8,$  the model  (\ref{reserves}),  (\ref{alcoupled1}), (\ref{alcoupled2}), (\ref{alcoupled3}), (\ref{logalCI}),   (\ref{rho})  reduces to the following (sub)-models:
\begin{equation}\label{reservesInt}
c_{t}^{i}=a_{t}^{i}-l_{t}^{i}, \quad   t\in [\tau_{1}^{j},\tau_{2}^{j}], \,  i=1,2,\ldots,N ,  \, \,  j=0,1,\ldots,8, 
\end{equation}
where the stochastic processes:
\begin{equation} \label{alcoupled1Int}
G_{t}^{i}=\ln(a_{t}^{i}), \quad H_{t}^{i}=\ln(l_{t}^{i}),  \quad    t\in [\tau_{1}^{j},\tau_{2}^{j}], \,  i=1,2,\ldots,N ,  \, \,  j=0,1,\ldots,8, 
\end{equation}
satisfy the following system of stochastic differential equations:
\begin{eqnarray}
&& d G_{t}^{i}=  \frac{\alpha_{t}}{N}\sum_{k=1}^{N} \left( G_{t}^{k} -G_{t}^{i}\right) dt  + d \ln(\varphi_{t}) + \sigma_{a}  dW_{t}^{i} , \nonumber\\
&&\hskip4truecm  \quad   t\in (\tau_{1}^{j},\tau_{2}^{j}], \,  i=1,2,\ldots,N ,  \, \,  j=0,1,\ldots,8,    \label{alcoupled2Int}\\
&& d H_{t}^{i}=  \frac{\gamma_{t}}{N}\sum_{k=1}^{N} \left( H_{t}^{k} -H_{t}^{i}\right) dt  +d \ln(\phi_{t}) + \sigma_{l}  dZ_{t}^{i} , \nonumber\\
&&\hskip4truecm  \quad   t\in (\tau_{1}^{j},\tau_{2}^{j}], \,  i=1,2,\ldots,N ,  \, \,  j=0,1,\ldots,8,    \label{alcoupled3Int}
\end{eqnarray}
with the  initial conditions:
\begin{eqnarray}
&&G_{\tau_{1}^{0}}^{i}=\ln(\tilde{a}_{0}),  \quad H_{\tau_{1}^{0}}^{i}=\ln(\tilde{l}_{0}), \quad   \,  i=1,2,\ldots,N,  \label{logalCIInt1}\\
&&
G_{\tau_{1}^{j}}^{i}= G_{\tau_{2}^{j-1}}^{i},    \quad H_{\tau_{1}^{j}}^{i}= H_{\tau_{2}^{j-1}}^{i}, 
 \quad   \,  i=1,2,\ldots,N ,  \, \,  j=1,2,\ldots,8,   \label{logalCIInt2}\
\end{eqnarray}
and the assumption:
\begin{eqnarray}\label{rhoInt}
&&\E(dW_{t}^{i} dW_{t}^{k})=\rho_{a}^{2} \, dt,\,   \, \, i\ne k, \qquad  \E(dZ_{t}^{i} dZ_{t}^{k})=\rho_{l}^{2} \, dt,\,\,   \, \, i\ne k, \, \,    \nonumber\\[3mm]
&&\E(dW_{t}^{i} dW_{t}^{i})=\E(dZ_{t}^{k} dZ_{t}^{k})=dt,   \qquad  
 \E(dW_{t}^{i} dZ_{t}^{k})=0, \, \,    \nonumber\\[3mm] 
 &&\hskip4truecm  t\in [\tau_{1}^{j},\tau_{2}^{j}], \quad i,k=1,2,\ldots,N,  \, \,  j=0,1,\ldots,8.  
\end{eqnarray}
For  $j=0,1,\ldots,8,$  the functions
  $\alpha_{t}$, $\gamma_{t}$,  $ t \in  [\tau_{1}^{j},\tau_{2}^{j}]$, in (\ref{alcoupled2Int}), (\ref{alcoupled3Int})  chosen   are obtained adapting formula       (\ref{alphagammaott})
 that is relative to  the    time interval $[0,T_{1}]$  to the time interval $ [\tau_{1}^{j},\tau_{2}^{j}]$.
In each time interval $ [\tau_{1}^{j},\tau_{2}^{j}]$, $j=0,1,\ldots,8,$   the probability  of    systemic risk   of    the corresponding  sub-model 
  (\ref{reservesInt}),  (\ref{alcoupled1Int}), (\ref{alcoupled2Int}), (\ref{alcoupled3Int}), (\ref{logalCIInt1}),   (\ref{logalCIInt2}),   (\ref{rhoInt})
      is evaluated using  statistical simulation starting from $10^{4}$  numerically  generated trajectories of  the corresponding sub-model   (\ref{reservesInt}),  (\ref{alcoupled1Int}), (\ref{alcoupled2Int}), (\ref{alcoupled3Int}), (\ref{logalCIInt1}),   (\ref{logalCIInt2}),   (\ref{rhoInt}). These trajectories are  obtained  by finite differences 
using   the explicit Euler method   with time step $\Delta t=10^{-4} $ to solve numerically the stochastic differential  equations   
 (\ref{alcoupled2Int}), (\ref{alcoupled3Int}) with the auxiliary conditions (\ref{logalCIInt1}),   (\ref{logalCIInt2}),   (\ref{rhoInt}).
   
In order to keep  the probability  of systemic risk   in each time  interval $ [\tau_{1}^{j},\tau_{2}^{j}]$, $j=0,1,\ldots,8,$     between the   thresholds  $S_{1}$ and $S_{2}$, we provide  to the monetary authority  a  pre-defined set    of functions  that can be used   to   push  the trajectories   of the assets and of the liabilities of  the $j$-th   sub-model     (\ref{reservesInt}),  (\ref{alcoupled1Int}), (\ref{alcoupled2Int}), (\ref{alcoupled3Int}), (\ref{logalCIInt1}),   (\ref{logalCIInt2}),   (\ref{rhoInt})  ``upward'' or  ``downward'',  or to  leave them  ``unchanged'', $j=0,1,\ldots,8$.
That is  for the assets we define the  functions: 
\begin{eqnarray} \label{possibilityAssets}
\nonumber A_{j,n_{a}}:  \quad \quad 
\varphi_{t}=\varphi_{t,j,n_{a}}=
\left\{%
\begin{array}{ll}
  \displaystyle \frac{n_{a}}{8} \,  (t-\tau_{1}^{j})+\varphi_{\tau_{1}^{j}} ,  & t\in [\tau_{1}^{j},\tau_{1}^{j} + \Delta \tau]  , \\[4mm]
  \displaystyle  \frac{n_{a}}{8}  \, \Delta \tau+\varphi_{\tau_{1}^{j}} ,  & t\in (\tau_{1}^{j} + \Delta \tau, \tau_{2}^{j}]  , 
\end{array}%
\right.\\[4mm]
 \quad j=0,1,\ldots,8, \, \, \, n_{a}=-8,-7,..,0,..,7,8,
 \end{eqnarray}
 similarly  for the liabilities we  define the    functions: 
\begin{eqnarray} \label{possibilityLiabilities}
\nonumber L_{j,n_{l}}:  \quad \quad 
\phi_{t}=\phi_{t,j,n_{l}}=
\left\{%
\begin{array}{ll}
  \displaystyle \frac{n_{l}}{8} \,  (t-\tau_{1}^{j})+\phi_{\tau_{1}^{j}} ,  & t\in [\tau_{1}^{j},\tau_{1}^{j} + \Delta \tau]  , \\[4mm]
  \displaystyle  \frac{n_{l}}{8}  \, \Delta \tau+\phi_{\tau_{1}^{j}} ,  & t\in (\tau_{1}^{j} + \Delta \tau, \tau_{2}^{j}]  , 
\end{array}%
\right.\\[4mm]
 \quad j=0,1,\ldots,8, \, \,  \, n_{l}=-8,-7,..,0,..,7,8,
 \end{eqnarray}
 and finally  based on  (\ref{possibilityAssets}), (\ref{possibilityLiabilities})  for the capital reserves we  define the    functions: 
 \begin{eqnarray} \label{possibility}
 P_{j,n_{a},n_{l}}: &&  
\xi_{t}=\xi_{t,j,n_{a},n_{l}}=\varphi_{t,j,n_{a}}-\phi_{t,j,n_{l}},  \quad t\in [\tau_{1}^{j}, \tau_{2}^{j}] , \nonumber\\[2mm]
&&  \quad \quad  j=0,1,\ldots,8, \, \, \, n_{a} \, n_{l}=-8,-7,..,0,..,7,8.
 \end{eqnarray}
Note that for $j=0,1,\ldots,8$  in (\ref{possibilityAssets})  when  $0<n_{a}\le 8$ (respectively $-8\le n_{a}<0$)  the   function $\varphi_{t,j,n_{a}}$,  is a  non decreasing  (respectively non increasing) piecewise linear function of $t$, while when $n_{a}=0$  the   function  $\varphi_{t,j,n_{a}}$  is a   constant. 
 Consequently for $j=0,1,\ldots,8,$ the choices  of   functions  $\varphi_{t,j,n_{a}}$  with   $0<n_{a}\le 8$ (respectively $-8\le n_{a}<0$)  in (\ref{possibilityAssets})    push  the trajectories   of the assets of  the $j$-th  sub-model    (\ref{reservesInt}),  (\ref{alcoupled1Int}), (\ref{alcoupled2Int}), (\ref{alcoupled3Int}), (\ref{logalCIInt1}),   (\ref{logalCIInt2}),   (\ref{rhoInt})   ``upward''   (respectively   ``downward''), while the  choice  $n_{a}=0$    in  (\ref{possibilityAssets})  leaves the trajectories  of the assets of the $j$-th sub-model   (\ref{reservesInt}),  (\ref{alcoupled1Int}), (\ref{alcoupled2Int}), (\ref{alcoupled3Int}), (\ref{logalCIInt1}),   (\ref{logalCIInt2}),   (\ref{rhoInt}) ``unchanged''.  Similar   statements  adapted to the circumstances hold  for the choices   $0< n_{l}\le 8$, $-8\le n_{l}<0$, $n_{l}=0$  of the functions $\phi_{t,j,n_{l}}$  in (\ref{possibilityLiabilities})  and for the trajectories   of the liabilities  of  the $j$-th  sub-model    (\ref{reservesInt}),  (\ref{alcoupled1Int}), (\ref{alcoupled2Int}), (\ref{alcoupled3Int}), (\ref{logalCIInt1}),   (\ref{logalCIInt2}),   (\ref{rhoInt}).  
Note that the implementation of {\it Strategy 1}, {\it  2},  {\it  3}  with the  choices made    in (\ref{possibilityAssets}), (\ref{possibilityLiabilities}) is only    illustrative, many   other choices of  the functions representing the  assets and  the liabilities of the  ``ideal bank''   are possible and lead to results analogous  to the ones discussed here.

To   measure the quality and  the cost of the systemic risk governance implemented in the experiments we   define   four  performance  indices.
Let
 \begin{eqnarray} \label{eta}
 \eta_{j}=\pr (SR_{[\tau_{1}^{j},\tau_{2}^{j}]}) ,  \quad   j=0,1,\ldots,8,
 \end{eqnarray}
 and let $ \underline{\eta}=
( \eta_{0}, \eta_{1},\ldots,\eta_{8}) \in \R^{9}$ be the  vector of the   systemic risk governance  procedure implemented in  the
 experiments. The systemic risk norm  ${\cal N }_{SR}$  is defined   as follows:
 \begin{eqnarray} \label{eff}
\displaystyle {\cal N }_{SR}=\left\| \underline{\eta}
\right\|_{2}  ,
 \end{eqnarray}
where $\|\underline{\eta}\|_{2}$ denotes the Euclidean norm of the  vector $ \underline{\eta}$.
The index ${\cal N }_{SR}$ is used to measure the quality of the systemic risk governance. Note that small  values of the index ${\cal N }_{SR}$  
correspond to high quality  systemic risk governance and that    in  the  numerical  experiments discussed here  when the goal of the governance (\ref{striscia}) is achieved in every one year time interval contained  in   $[0,T_{2}]$ we have 
$ 3 S_{1} \le {\cal N }_{SR} \le 3 S_{2}$.

The indices  ${\cal C }_{SR}^{c}$,  ${\cal C }_{SR}^{\alpha}$, ${\cal C }_{SR}^{\gamma}$  are used to  measure   the cost of the  systemic risk governance. 
The first index ${\cal C }_{SR}^{c}$  measures  the ``cost  associated      to the choice of the assets and of the liabilities''  of the  ``ideal bank''   defined   in (\ref{possibilityAssets}) and (\ref{possibilityLiabilities}), while   the indices  ${\cal C }_{SR}^{\alpha}$, ${\cal C }_{SR}^{\gamma}$  measure ``the cost  associated to the 
choice  of the functions  $\alpha_{t}$,  $\gamma_{t}$,  $ t \in [0,T_{2}]$,  that regulate  the cooperation  mechanism among banks.
More specifically  in each one year period  considered in the governance procedure   we  define as  cost associated to the choice of  the  assets and of the liabilities  of the  ``ideal bank'' 
the  absolute value of the angular coefficient of the linear part of the  piecewise linear functions listed in  (\ref{possibilityAssets}), (\ref{possibilityLiabilities}).
In this way in the period $ [\tau_{1}^{j},\tau_{2}^{j}]$  
the cost of  choosing $A_{j,n_{a}}$, defined in   (\ref{possibilityAssets}),  is  $\displaystyle \frac{|n_{a}|}{8}$, $n_{a}=-8,-7,..,0,..,7,8,$ and similarly   
the cost of choosing $L_{j,n_{l}}$, defined  in  (\ref{possibilityLiabilities}),  is  $\displaystyle \frac{|n_{l}|}{8}$, $n_{l}=-8,-7,..,0,..,7,8$.  
Finally the cost  of choosing
$ P_{j,n_{a},n_{l}} $  given   in (\ref{possibility}),  is  defined as $\displaystyle  \frac{|n_{a}|}{8}+ \frac{|n_{l}|}{8}$, $n_{a}=-8,-7,..,0,..,7,8$,  $n_{l}=-8,-7,..,0,..,7,8$.
The total  cost measured by the index   ${\cal C }_{SR}^{c}$  of    the  systemic risk governance procedure  defined above  is given by the sum over $j$   of the cost of  the  
trajectories  $ P_{j,n_{a},n_{l}}$, $j=0,1,\ldots,8,$  used in the  procedure.
The indices   ${\cal C }_{SR}^{\alpha}$, ${\cal C }_{SR}^{\gamma}$ are given,  respectively, by the  sum of the means  of $\alpha_{t}$,  $\gamma_{t}$
in the  time intervals  $ [\tau_{1}^{j},\tau_{2}^{j}]$, $j=0,1,\ldots,8,$  
used in the systemic risk governance procedure, that is,  recalling equations  (\ref{alcoupled2Int}), (\ref{alcoupled3Int}) and defining:
 \begin{eqnarray} 
&&\bar{\alpha}_{j}=\frac{1}{\Delta \tau}   \int_{\tau_{1}^{j}}^{\tau_{2}^{j}}\alpha_{t} \ dt,        \quad t\in [\tau_{1}^{j},\tau_{2}^{j}], \quad  j=0,1,\ldots,8, \\
&&\bar{\gamma}_{j}=\frac{1}{\Delta \tau}   \int_{\tau_{1}^{j}}^{\tau_{2}^{j}}\gamma_{t} \ dt, , \quad t\in [\tau_{1}^{j},\tau_{2}^{j}], \quad  j=0,1,\ldots,8,
 \end{eqnarray}
we have:
 \begin{eqnarray} \label{inidicicosto}
{\cal C }_{SR}^{\alpha}=\sum_{j=0}^{8}  \bar{\alpha}_{j},  \quad \quad  {\cal C }_{SR}^{\gamma}=\sum_{j=0}^{8}   \bar{\gamma}_{j}.
 \end{eqnarray}

In   the numerical experiments the indices ${\cal N }_{SR}$,  ${\cal C }_{SR}^{c}$,  ${\cal C }_{SR}^{\alpha}$, ${\cal C }_{SR}^{\gamma}$
    change  significantly  depending from  the  circumstances (i.e. presence  or absence of volatility and correlation   shocks)  faced  during   the two years period of the  systemic risk governance procedure.
   Moreover covering  the  entire history  of the governance, that  is covering  a two year  governance period made of   nine quarterly  decisions, 
    the indices  defined above measure only  a  ``overall'' quality and cost of the  systemic risk governance procedure.

Table \ref{Tab1} shows  the  numerical  results obtained in the  systemic risk governance  of model (\ref{reservesInt}),  (\ref{alcoupled1Int}), (\ref{alcoupled2Int}), (\ref{alcoupled3Int}), (\ref{logalCIInt1}),   (\ref{logalCIInt2}),   (\ref{rhoInt}). 
  In the experiments  presented the monetary authority  pursues  the goal of keeping  the probability of  systemic risk in the next year between the thresholds  $S_{1}=0.01$ and $S_{2}=0.05$  implementing the actions associated to  {\it Strategy 1},  {\it 2},  {\it  3} 
through the  choice of the functions  $A_{j,n_{a}}$, $L_{j,n_{l}}$, $ P_{j,n_{a},n_{l}} $,  $j=0,1,\ldots,8$,  $n_{a}=-8,-7,..,0,..,7,8$,  $n_{l}=-8,-7,..,0,..,7,8$, defined respectively in 
 (\ref{possibilityAssets}),  (\ref{possibilityLiabilities}),   (\ref{possibility}).
More in detail,     for  $ j=0,1,\ldots,8, $  the monetary authority runs through the  possible choices  of  the functions  listed in  (\ref{possibilityAssets}), (\ref{possibilityLiabilities}),   (\ref{possibility})     in their natural order  according to    {\it Strategies 1}, {\it 2}, {\it  3} starting from the choice    $ A_{j,0}$, $ L_{j,0}$, $  P_{j,0,0}=A_{j,0}- L_{j,0}$     and evaluates using statistical simulation  the probability of systemic risk in the next year associated to each choice of the previous  functions considered. The first choice  encountered   that  gives a  probability  of systemic risk in the next year that satisfies (\ref{striscia})  is chosen as systemic risk governance decision. 
The choice of the functions $\alpha_{t}$,  $\gamma_{t}$ corresponding to the previous choices of the  functions   $\varphi_{t}$,   $\phi_{t} $,   is done 
adapting   (\ref{alphagammaott})  to the circumstances. 
If none of the functions  listed in  (\ref{possibilityAssets}), (\ref{possibilityLiabilities}),   (\ref{possibility})   gives a  probability  of systemic risk in the next year that satisfies (\ref{striscia}) the governance procedure  is not able to  reach its goal   in the  time interval  considered,  in this case  the governance procedure  takes the best choice available in  (\ref{possibilityAssets}), (\ref{possibilityLiabilities}),   (\ref{possibility})   and tries to reach its  goal  in the successive time interval.

Note that  when the correlation coefficients $ \rho_{a}^{2}$ and/or $\rho_{l}^{2}$ increase  the ``swarming'' effect induced by  the  cooperation mechanism among banks   in  (\ref{alcoupled2}) and/or  (\ref{alcoupled3})  decreases. Recall that in the extreme case of
  $\rho_{a}^{2} = 1$,  $ \rho_{l}^{2}  = 1$   the  cooperation mechanism  has  no effect anymore.
Therefore  when  $\rho_{a}^{2} = 1$,  $ \rho_{l}^{2}  = 1$    to govern the  systemic risk probability it is  only possible  is   to increase the capital reserves   of the ``ideal bank''. 

Let $\sigma_{4,t}$, $t \in  [0,T_{2}]$, $T_{2}=3$,    be defined as follows:
\begin{equation}  \label{sigma4t}
\sigma_{4,t}=
\left\{%
\begin{array}{ll}
  \displaystyle 0.3 ,  & t\in [0,0.2]  , \\
  \displaystyle  0.8 ,  & t\in (0.2, 0.5],\\ 
   \displaystyle  0.3 ,  & t\in (0.5,3] .
\end{array}%
\right.
\end{equation}
 In the  experiments of Table \ref{Tab1}  to  model positive  shocks acting on the  assets or on  the  liabilities of the banks we consider  the choices    $\sigma_{a}=\sigma_{4,t}$, $t \in  [0,T_{2}]$, $T_{2}=3$,    or     $\sigma_{l}=\sigma_{4,t}$, $t \in  [0,T_{2}]$, $T_{2}=3$.
 Moreover    let   $\rho_{1,t}$, $t \in  [0,T_{2}]$, $T_{2}=3$,    be defined  as follows:
\begin{equation}  \label{rho1t}
\rho=\rho_{1,t}=
\left\{%
\begin{array}{ll}
  \displaystyle 0 ,  & t\in [0,0.2]  , \\
  \displaystyle  0.5 ,  & t\in (0.2, 0.5],\\ 
   \displaystyle  0,  & t\in (0.5,3] .
\end{array}%
\right.
\end{equation}
The ``collective''  behaviour of the banks in presence  of shocks  is  modeled assuming a positive correlation in the noise terms of the  assets or  of the  liabilities equations  of  (\ref{reservesInt}),  (\ref{alcoupled1Int}), (\ref{alcoupled2Int}), (\ref{alcoupled3Int}), (\ref{logalCIInt1}),   (\ref{logalCIInt2}),   (\ref{rhoInt}). In particular in some experiments we consider    the choices   $\rho_{a}=\rho_{1,t}$, 
$t \in  [0,T_{2}]$, $T_{2}=3$,      or      $\rho_{l}=\rho_{1,t}$, $t \in  [0,T_{2}]$, $T_{2}=3$.  

Note that the function $\sigma_{4,t}$, $t \in  [0,T_{2}]$, $T_{2}=3$,  defined   in (\ref{sigma4t}) and  the function  $\rho_{1,t}$, $t \in  [0,T_{2}]$, $T_{2}=3$, defined  in (\ref{rho1t}) ``jump together''  at time $t=0.2$ and $t=0.5$.

The remaining  parameters  of the model  used in the experiments reported in   Table \ref{Tab1}    are:  $\mu_{a}=0.1$, $\mu_{l}=0.1$, 
$\tilde{a}_{0}=\varphi_{0}=0.6$, $\tilde{l}_{0}=\phi_{0}=0.4$, $\lambda_{i}=0.1$, $i=1,2,3,4$.  
Note that the previous choices   guarantee $\xi_{t}=\varphi_{t}- \phi_{t}>0$, $t \in  [0,T_{2}]$, $T_{2}=3$.

Table \ref{Tab1}  shows the values of  the indices ${\cal N }_{SR}$,  ${\cal C }_{SR}^{c}$,  ${\cal C }_{SR}^{\alpha}$, ${\cal C }_{SR}^{\gamma}$ obtained in  the   experiments. 
  In the seventh column  of Table \ref{Tab1},  next to the value of ${\cal N }_{SR}$,  it is shown, within brackets,  the value of   ${\cal N }_{SR}$  obtained   in absence of governance. 
 In    absence of governance we choose  
 $\mu_{a}=0.1$, $\mu_{l}=0.1$,  $\alpha_{t}=10$,  $\gamma_{t}=10$, $t\in [0,T_{2}]$, $T_{2}=3$, $\tilde{a}_{0}=\varphi_{0}=0.6$, $\tilde{l}_{0}=\phi_{0}=0.4$,  and we evaluate
the probability of   systemic risk in the next year   at  the beginning of each  quarter. 
These     choices  of the parameter values   guarantee that   the probability  of systemic risk   in the next year  at time $t=0$  is between the    thresholds $S_{1}=0.01$ and $S_{2}=0.05$. 
Note that with the previous choices  in absence of governance  the values of the  indices  ${\cal C }_{SR}^{c}$,  ${\cal C }_{SR}^{\alpha}$, ${\cal C }_{SR}^{\gamma}$   are always equal to, respectively, 0, 90, 90.  When necessary   the values of the  last three columns of  Table \ref{Tab1}  obtained  in presence  of governance 
 may be  compared with values of the indices    ${\cal C }_{SR}^{c}=0$,  ${\cal C }_{SR}^{\alpha}=90$, ${\cal C }_{SR}^{\gamma}=90$  that correspond to the absence of governance.
  
A first overview  of Table \ref{Tab1}  shows that from the  systemic risk governance
point of view  the performance  of   Strategies 
{\it 1a, 2a, 3} versus    Strategies {\it 1a, 2b, 3}   and of   Strategies {\it 1a, 2b, 3} versus    Strategies  {\it 1b, 2b, 3} 
is  approximately the same. 
 
Experiments 1 and 2 of   Table \ref{Tab1} show that when the volatilities $\sigma_{a}$, $\sigma_{l}$ are constants  and there are no correlation  shocks   in the time interval $ [0,T_{2}]$, $T_{2}=3$,
the presence or  the absence of governance does not make a significant difference provided that,  in absence of governance at time $t=0$,   a good choice of the
assets and of the liabilities of the ``ideal bank''  in the one year period   beginning at  time $t=0$ and of the constant   values of
 $\alpha_{t}$,   $\gamma_{t}$, $t\in [0,T_{2}]$, $T_{2}=3$, is done.  
That is,  when the  volatilities $\sigma_{a}$, $\sigma_{l}$ are constants  and there are no correlation  shocks  
 in the time interval $ [0,T_{2}]$, $T_{2}=3$,  the systemic risk  governance     is substantially reduced to the choice of   the assets and  of the  liabilities of the ``ideal bank''  at the beginning of the time interval considered for the governance experiment, that is at time $t=0$ and  of the constant   values of
the functions  $\alpha_{t}$,   $\gamma_{t}$, $t\in [0,T_{2}]$, $T_{2}=3$.  When this choice is done correctly   continuing with  
 assets and  liabilities functions of the ``ideal bank''  constants   or with small variations of it   in each successive time interval   $ [\tau_{1}^{j},\tau_{2}^{j}]$,  $j=1,2,\ldots,8$,   is  sufficient to  keep   the probability  of systemic risk  in the next year  between the   given thresholds.
 The functions  $\alpha_{t}$,   $\gamma_{t}$, $t\in [0,T_{2}]$, $T_{2}=3$, is done according to the rules established in Section \ref{sec5} in the study of the control problem for the   pseudo mean field approximation  of the   banking system.
  In this case the possibly  positive value of ${\cal C }_{SR}^{c}$  in presence of governance  (compared with  ${\cal C }_{SR}^{c}=0$ of the  absence of governance)   is  certainly compensated by the smaller values of the indices ${\cal C }_{SR}^{\alpha}$, ${\cal C }_{SR}^{\gamma}$  with respect to the values corresponding to the  absence of governance (i.e. ${\cal C }_{SR}^{\alpha}$=90, ${\cal C }_{SR}^{\gamma}=90$).
 Moreover note that in these cases  we have  $ 3 S_{1} \le {\cal N }_{SR} \le 3 S_{2}$.

Note  that  in Experiment 1  the choice  of the functions $\alpha_{t}=10$,  $\gamma_{t}=10$, $t\in [0,T_{2}]$, $T_{2}=3$, made in absence of governance   together with the choices of the other parameters of the problem guarantees 
that   the probability  of systemic risk   in the next year  at time $t=0$  is between the    thresholds $S_{1}=0.01$ and $S_{2}=0.05$ and   gives values of 
${\cal N }_{SR}$ in presence and in absence of  governance of the  same order of magnitude.  
To have an idea  of the consequences of changing the values  $\alpha_{t}=10$,  $\gamma_{t}=10$, $t\in [0,T_{2}]$, $T_{2}=3$, let us mention  that 
if in  Experiments 1   we fix  $\alpha_{t}=1$,  $\gamma_{t}=1$, $t\in [0,T_{2}]$, $T_{2}=3$,  leaving all the remaining parameters unchanged,  in absence of governance we have ${\cal C }_{SR}^{c}=0$,  ${\cal C }_{SR}^{\alpha}=9$, ${\cal C }_{SR}^{\gamma}=9$, but we have ${\cal N }_{SR}=0.30$.

As expected the  comparison between  Experiments 1 and 3   of   Table \ref{Tab1} shows  that    the governance of the systemic risk in presence of  volatility shocks  is more demanding than the governance in absence of  shocks. 
This can be seen comparing  the cost indices ${\cal C }_{SR}^{c}$,  ${\cal C }_{SR}^{\alpha}$, ${\cal C }_{SR}^{\gamma}$ and the quality index 
${\cal N }_{SR}$  of Experiment 1 and 3. 
Furthermore things become  worse when    together with  a volatility shock also    a
  correlation  shock acts on the    liabilities equations  of the banking system. 
 This last fact  can be seen  comparing  the performance indices of  Experiments 1, 3 and 4  of   Table \ref{Tab1}.
 In particular   the comparison of the performance indices of   Experiments  3 and 4  shows that the values of the cost indices  ${\cal C }_{SR}^{c}$,  ${\cal C }_{SR}^{\alpha}$, ${\cal C }_{SR}^{\gamma}$ increase significantly going from Experiment  3 to Experiment 4
 despite the fact that the index ${\cal N }_{SR}$ signals that  the quality of the governance is decreasing. 
 In fact in Experiment 3 and 4 we have ${\cal N }_{SR}>3 S_{2}$, this means that during the two year period studied has not been always possible to satisfy       (\ref{striscia}).
 Note that the index ${\cal N }_{SR}$  in Experiment 4 is greater  than in Experiment 3, 
 this shows  that the ``collective''  behaviour   of the banks
 induced by the non zero correlation $\rho_{a}$  makes the governance more difficult. 

Similar observations   can be made  when  a volatility shocks  acts on  the   liabilities   of the banks  (compare, for example,  Experiments  1   and 5 of   Table \ref{Tab1}) and when 
 positive correlation  is present in the noise terms of  the  liabilities equations $\rho_{l}$ (compare, for example,   Experiments  5 and 6).

 Moreover   note that   in  Experiments 3, 4, 5, 6    the values of the index ${\cal N }_{SR} $ is  always greater than  $3 S_{2}=0.15$. This is due to the fact that   when the governance  faces  the  shock for the first time it is unable to reach its goal of having  the  probability of  systemic risk  in the next year inside the assigned thresholds.

We  conclude that     in   the Experiments presented    in Table \ref{Tab1}   the   systemic risk  governance procedure  proposed   is able  to reach its goal, that is is able  to  keep 
  the probability   of    systemic risk   in the next year between the    assigned thresholds  
  at a reasonable cost.
  




\vskip1truecm

 \newpage
 \begin{table}[h]
 \small
\centerline{
\begin{tabular}{|c|c|c|c|c|c||c||c|c |c| c |}
\hline
{\it Experi-} & $\sigma_{a}$  & $\sigma_{l}$ &$\rho_{a}$  & $\rho_{l}$  & {\it Strategies} & ${\cal N }_{SR}$ ( ${\cal N }_{SR}$ no gov) &   ${\cal C }_{SR}^{c}$       & $    {\cal C }_{SR}^{\alpha}$       & $ {\cal C }_{SR}^{\gamma}  $ \\
{\it ment} &   &  &  &   &  &   &          &       & \\
\hline
1  & $ 0.3$ & $ 0.3$ & $0$ & $0$ &{\it 1a, 2a, 3}   & 0.08 (0.06)  &  0.28  &  4.17   &   4.17  \\[3mm]
&&&&&{\it 1a, 2b, 3}   &  0.08 (0.06)  &  0.28      &  4.17     &  4.17  \\
\hline
2  & $ 0.3$ & $ 0.3$ &  $\rho_{1,t}$ &$0$ &{\it 1a, 2a, 3}   & 0.07 (0.13)   & 0.33    &  5.59    &     4.59  \\[3mm]
&&&&&{\it 1a, 2b, 3}     & 0.06 (0.13)   & 0.33      & 5.59     &    4.59  \\
\hline
\hline
3  &  $ \sigma_{4,t}$ & $ 0.3$ & $0$ & $0$ &{\it 1a, 2a, 3}   &0.17 (0.71)   &  1.26     & 12.91       &  12.91  \\[3mm]
&&&&&{\it 1a, 2b, 3}   &  0.19 (0.71)  &  1.21     &  11.77       &   11.77 \\
\hline
4    &  $ \sigma_{4,t}$ & $0.3 $& $\rho_{1,t}$ & $0$ &{\it 1a, 2a, 3}   & 0.26 (0.77)  & 1.53      & 21.99       &  15.72  \\[3mm]
&&&&&{\it 1a, 2b, 3}   & 0.25 (0.77)  & 1.34     & 19.33         & 13.07   \\
\hline
\hline
5  &  $ 0.3$ & $ \sigma_{4,t}$ & $0$  & $0$ &{\it 1a, 2b, 3}   &  0.20 (0.72) &  1.21     &  11.77      & 11.77  \\[3mm]
&&&&&{\it 1b, 2b, 3}  & 0.18 (0.72)   &   1.11    &   10.41      &   10.41 \\
\hline
6  &  $ 0.3$ & $ \sigma_{4,t}$ & $0$ & $\rho_{1,t}$ &{\it 1a, 2b, 3}   & 0.27 (0.77)   &   1.36    &  13.45     &  19.71  \\[3mm]
&&&&&{\it 1b, 2b, 3}   & 0.50 (0.77)  &    2.01      & 24.45     &    30.69\\
\hline
 \end{tabular}
 }\vspace{-0.5cm}
 \bigskip\caption{Numerical experiments with $\mu_{a}=\mu_{l}=0.1$,  $\lambda_{i}=0.1$, $i=1,2,3,4$, $\varphi_{0}=0.6$,  $\phi_{0}  =0.2$, $S_{1}=0.01$, $S_{2}=0.05$. }\label{Tab1}
\end{table}

\end{document}